\documentclass[12pt,english]{article}
\usepackage[breaklinks,hypertexnames=false]{hyperref}
\hypersetup{colorlinks=true,linkcolor=blue,urlcolor=blue,citecolor=blue}
\usepackage{amsthm}
\usepackage{amsmath}
\numberwithin{equation}{section}
\usepackage{amssymb}
\usepackage{esint}
\usepackage[authoryear]{natbib}
\usepackage{listings}
\usepackage{inconsolata}

\usepackage{subcaption} 

\usepackage{appendix}
\usepackage[english]{babel}
\usepackage{enumitem}
\usepackage{booktabs}
\usepackage{multirow}
\usepackage{natbib}
\usepackage{textcomp,mathcomp}
\usepackage{adjustbox}
\usepackage{dcolumn}
\usepackage{booktabs}
\usepackage{mathabx} 
\usepackage{float}
\usepackage{lmodern}    
\usepackage[T1]{fontenc}

\makeatletter
\usepackage{graphicx}
\usepackage{lscape}
\usepackage{rotating}
\usepackage[english]{babel}
\usepackage{latexsym}	
\usepackage{bbm}
\usepackage{tikz}
\usetikzlibrary{arrows,shapes,trees,positioning}
\usepackage{seqsplit}
\usepackage{natbib}
\usepackage{pgfplots}
\usepgfplotslibrary{groupplots}

\hypersetup{
	colorlinks=true,
	citecolor=blue
}

\makeatother
\newtheorem{theorem}{Theorem}

\newtheorem{remark}[theorem]{Remark}

\newcommand{\defeq}{\vcentcolon=}
\renewcommand{\P}{\mathcal{P}}

\usepackage{todonotes}
\usepackage{tikz}

\defcitealias{arellano2017earnings}{ABB}

\usepackage{babel}

\usepackage{ragged2e}

\newcommand{\figurenote}[1]{%
  \captionsetup{justification=justified, singlelinecheck=false}%
  \vspace{0.3em}%
  \begin{minipage}{0.9\textwidth}
    \setlength{\parindent}{0.0em}%
    \footnotesize
    \justifying
    \noindent \textit{Note:} #1
  \end{minipage}%
}

\usepgfplotslibrary{fillbetween}
\usepackage{standalone}
\pgfplotsset{compat=1.17}

\usepackage{pgfplots}
\usepgfplotslibrary{groupplots}     
\usepgfplotslibrary{fillbetween}    
\usetikzlibrary{arrows.meta}        
\usepackage{amsmath}
\usepackage{booktabs}               
\usepackage{float}                  
\usepackage{graphicx}               
\usepackage{subcaption}             

\definecolor{clogp}{HTML}{1B3A6B}
\definecolor{cnon}{HTML}{007373}
\definecolor{camo}{HTML}{D4683A}
\definecolor{cfam}{HTML}{B0B3B8}
\definecolor{camg}{HTML}{C2AEE6}
\definecolor{cgrid}{HTML}{E6E8EB}

\begin{document}
	
\title{Indirect Variational Inference:\\ Applications to Earnings Dynamics\thanks{A previous version of the paper was circulated with the title ``How Variational Are Earnings Dynamics?'' We thank the Editor Joe Altonji, two anonymous referees, and our discussant Manuel Arellano, as well as the audience of the 2025 Yale conference in honor of Costas Meghir, for helpful comments. This research was supported in part by the Pythia computing cluster at the University of Chicago Booth School of Business, which is funded by the Office of the Dean.}}
	\author{Neele Balke \\ 
    University of Chicago \and St\'{e}phane Bonhomme \\ University of Chicago \and Thibaut Lamadon \\ University of Chicago }
	\date{\today
	}
	\maketitle

	\vskip 1cm
	
	\begin{abstract}

Latent-variable models are central to economics but often entail intractable integration. Variational inference (VI), widely used in machine learning, turns this integration into tractable, differentiable optimization by replacing the likelihood with a variational objective. However, guarantees of recovering the true parameters remain limited when the variational family is insufficiently flexible—a key obstacle to the adoption of VI in economics. We first evaluate VI in models of earnings dynamics and show that the choice of variational posterior is crucial. We then introduce \textbf{indirect variational inference (IVI)}, which treats VI as an auxiliary model and corrects the bias induced by the variational approximation. IVI retains much of VI’s tractability because it does not require computing the likelihood. We apply these methods to models allowing for nonlinear persistence, non-Gaussian and serially correlated transitory shocks, and latent heterogeneity. Across simulated and empirical applications, flexible variational families combined with IVI deliver reliable estimates.

		\bigskip
		
		
	\end{abstract}

	\clearpage
	
	\global\long\def\ind{\mathbb{1}}
	\global\long\def\d{\mathrm{d}}
	\global\long\def\t{\intercal}
	\global\long\def\RR{\mathbb{R}}
	\global\long\def\defeq{:=}

\section{Introduction}

Understanding individual earnings dynamics is central to many questions in economics, from the design of social insurance to the study of consumption, saving, and inequality. A growing literature shows that earnings risk is not well described by linear Gaussian processes alone: nonlinear persistence, asymmetric shocks, and non-Gaussian innovations shape the evolution of earnings over the life cycle (e.g., \citealp{arellano2014uncertainty}, \citealp{ArellanoBlundellBonhomme2017}, \citealp{guvenen2021data}). These features motivate dynamic, nonlinear latent-variable models that decompose earnings into persistent and transitory components. Yet such models are difficult to estimate because the likelihood requires integrating over latent earnings histories and typically has no closed-form expression. Existing likelihood-based or simulation-based approaches can be delicate to tune and become increasingly costly as the time dimension or the model complexity grows.

This paper studies the use of \emph{variational inference} (VI) for estimating nonlinear latent-variable models of earnings dynamics. Variational inference replaces the likelihood with a tractable objective, the Evidence Lower Bound (ELBO), obtained by approximating the posterior distribution of the latent variables with a chosen family of variational posterior densities (e.g., \citealp{jordan1999introduction}, \citealp{kingma2014auto}; see \citealp{BleiKucukelbirMcAuliffe2017} for a survey). This turns repeated integration over latent trajectories into a differentiable optimization problem that is naturally compatible with automatic differentiation and stochastic gradient methods. However, the same approximation that makes VI fast and scalable to large panel data sets can also bias parameter estimates. Guarantees of recovering the true parameters remain limited when the variational family is insufficiently flexible, making the design of the variational posterior central to the reliability of the method.

To correct for the biases induced by restrictive variational approximations we introduce \emph{indirect variational inference} (IVI), which treats VI as an auxiliary model to perform indirect inference (\citealp{smith1993estimating}, \citealp{gourieroux1993indirect}, \citealp{gallant1996moments}). While variational inference is consistent for a pseudo-true value that generally differs from the truth, indirect variational inference is consistent and asymptotically normal for the true value, while retaining an important advantage of VI: it does not require computing or approximating the likelihood. In IVI, model and auxiliary parameters coincide, so the indirect inference problem is just-identified (as in \citealp{zhang2022just}). We propose two implementations, based on gradient descent and fixed-point iteration, respectively.

Before applying the method to PSID data, we evaluate VI and IVI on simulated panels from a sequence of earnings dynamics models of increasing complexity. The benchmark is the canonical persistent-transitory model, in which persistent earnings follow a linear Gaussian autoregression and transitory shocks are i.i.d. Gaussian. We then enrich this specification to allow for nonlinear persistence, state-dependent volatility, non-Gaussian and serially correlated transitory shocks, and time-invariant latent heterogeneity. Across these models, we approximate the posterior distribution of the latent components using Gaussian variational families, varying the restrictions imposed on their covariance structure.

In the linear Gaussian benchmark, VI recovers the model parameters well. This is expected since the true posterior is Gaussian and belongs to the class of variational families we consider. However, we find that correctly capturing the covariance structure of the variational posterior is crucial: substantial biases arise under a \emph{mean-field} approximation, a popular specification in machine learning that imposes independence across latent states over time. These biases are not necessarily an obstacle to parameter recovery when VI is used as an auxiliary estimator, since we find that IVI corrects much of the discrepancy between the VI estimates and the true parameter values.

We then turn to nonlinear and non-Gaussian specifications. Variational posteriors based on flexible Gaussian variational families recover the main features of the persistent component well, including nonlinear conditional means and state-dependent conditional variances. However, our VI estimates remain biased for other parameters, such as higher moments of transitory shocks, latent heterogeneity, or serial correlation in the transitory component. Across specifications, mean-field approximations perform poorly, while unrestricted Gaussian variational posteriors and structured variants that exploit the dynamic structure of the model perform substantially better. IVI further corrects the remaining biases and delivers estimates close to the true parameter values, including in longer panels where the scalability of VI is especially valuable.

We apply the method to annual data from the Panel Study of Income Dynamics (PSID) for the years 1980--1989. We estimate a flexible earnings dynamics model that combines nonlinear persistence, an MA(1) transitory component, and individual-specific transitory variances.  As shown in \citet{almuzara2020heterogeneity}, heterogeneity in transitory variances can potentially explain some of the observed nonlinearity in earnings, so accounting for heterogeneity and nonlinearity in the same model is empirically important. The comparison between VI and IVI then allows us to assess which empirical findings are robust to correcting for variational approximation bias. We find that the conditional mean of the persistent earnings component is approximately linear, with average persistence close to unity, while the conditional variance is nonlinear, displaying a U-shape across the distribution. We also uncover evidence of individual heterogeneity in transitory variances, and of serial correlation in transitory shocks. As in \citet{ArellanoBlundellBonhomme2017}, we find persistence to be nonlinear, in the sense that it is lower when high-income households experience negative shocks or low-income households experience positive shocks.

Overall, our analysis shows that variational approximation error can lead to substantial bias, but that this is not an inherent obstacle to reliable estimation. Two complementary strategies are important. The first is to rely on a sufficiently flexible variational family, since specifications such as mean-field can be too restrictive. The second is the use of the variational specification as an auxiliary model through indirect variational inference. The encouraging evidence we present motivates importing and improving these methods for their use in dynamic economic settings.

\paragraph{Literature. } 
A rich empirical literature has studied the dynamics of individual earnings, documenting important linear and nonlinear dynamic features. Key contributions include \citet{LillardWillis1978} and \citet{AbowdCard1989}, among many others. Early work modeled earnings as the sum of persistent and transitory components, typically assuming linear Gaussian processes. However, subsequent empirical studies have shown that these assumptions are too restrictive. \citet{MeghirPistaferri2004} develop econometric methods to separate transitory and permanent shocks to income using both earnings and consumption data, uncovering heterogeneity in variances across individuals. The broader literature, as reviewed by \citet{MeghirPistaferri2011}, emphasizes the importance of nonlinearities, including age-dependent volatility, state-dependent persistence, and heteroskedasticity as a source of earnings risk.

Recent contributions use nonlinear state-space models to allow for richer dynamics. \citet{ArellanoBlundellBonhomme2017} propose a quantile-based panel framework in which individual latent states evolve nonlinearly. This setup reveals that persistence is nonlinear, showing that shocks have different effects depending on past earnings. \citet{DeNardiFella2020} compare linear and nonlinear earnings processes in a structural life-cycle model and show that models with skewness, kurtosis, and state-dependent variances produce markedly different predictions for consumption inequality and self-insurance behavior than their linear Gaussian counterparts. \citet{braxton2024changing} develop a generalized Kalman filter and document how income risk varies along the skill distribution.


While variational inference has been successfully applied to many fields, including text analysis (\citealp{blei2003latent}) and computer vision (\citealp{kingma2014auto}), applications to economics are still relatively limited. \citet{ChanYu2022} develop variational methods for large Bayesian VARs with stochastic volatility. \citet{LoaizaMayaNibbering2023} apply variational inference to structural discrete choice models. \citet{mele2023approximate} and \citet{bonhomme2021teams} use variational inference to estimate network formation and team production models, respectively. \citet{kitagawa2023individualized} characterize the performance of a variational approximation to the optimization problem of individualized treatment assignment. 

In this paper, we focus on the empirical performance of variational inference on simulated and real data. There is also a literature on theoretical properties. A key question is whether parameters that maximize the evidence lower bound are asymptotically consistent for the true parameters of the model. \citet{bickel2013asymptotic} provide results for stochastic blockmodels, \citet{westling2019beyond} study mixture models, and \citet{katsevich2024approximation} study Gaussian variational inference when the true posterior is approximately Gaussian. \citet{wang2019frequentist} and \citet{wang2019variational} study frequentist properties of variational Bayes. \citet{wu2026extending} propose an entropic-regularized modification of mean-field variational inference. \citet{medina2022robustness} study the use of $\alpha$-posteriors and their variational approximations in the presence of model misspecification.

The remainder of the paper is organized as follows. In Section \ref{sec:model} we introduce a nonlinear latent-variable model of earnings dynamics, and in Section \ref{sec:VI-intro} we describe how variational inference can serve as a practical alternative to likelihood-based methods in this context. We introduce indirect variational inference in Section \ref{sec:IVI}. We then present simulation evidence based on a linear Gaussian data-generating process (DGP) in Section \ref{sec:benchmark}, and based on a nonlinear DGP in Section \ref{sec:nonlinear-model}. In Section \ref{sec:MA1} we show how to extend the approach to allow for time-invariant latent heterogeneity and serially correlated transitory shocks. In Section \ref{sec:PSID} we present an empirical application to the PSID. Finally, we conclude in Section \ref{sec:conclusion}. Additional details are provided in the appendix, and computer codes will be available online.

\section{A Nonlinear Model of Earnings Dynamics}\label{sec:model}

In this section, we introduce a nonlinear latent-variable model for individual earnings dynamics. The model has a hidden Markov structure with two latent components: persistent and transitory. Let \( y_t \) denote the log earnings of an individual (or household) at time \( t \), and let \( z_t \) denote a latent persistent component that evolves according to a state-dependent stochastic process. The observed data, for a given individual, are generated by:
\begin{align}
y_t &= z_t + e_t, \label{eq:mod_1}\\
z_t &= \mu(z_{t-1}) + \sigma(z_{t-1}) u_t, \label{eq:mod_2}\\
z_1 &\sim f_\alpha, \quad u_t \sim \mathcal{N}(0,1), \quad e_t \sim \psi_\gamma,\label{eq:mod_3}
\end{align}
where $u_t$ and $e_t$ are independent at all lags and independent of the initial $z_1$, and periods range from $1$ to $T$. While the latent components $z_{1:T}$ and $e_{1:T}$ are individual-specific, the parameters $\mu(\cdot)$, $\sigma(\cdot)$, $\alpha$, and $\gamma$ are common across individuals.

In this formulation, \( z_t \) represents the persistent component of individual earnings, while \( e_t \) are transitory shocks. The innovations \( u_t \) are standard normal, scaled by a state-dependent volatility function \( \sigma(z_{t-1}) \), and propagated forward through a conditional mean function \( \mu(z_{t-1}) \). Both \( \mu \) and \( \sigma \) are specified as flexible nonlinear functions, so that the model nests the standard linear earnings process but also allows for nonlinear and state-dependent dynamics. The initial latent state \( z_1 \) is drawn from a distribution \( f_\alpha \), which can itself be parameterized using flexible location-scale families to capture cross-sectional heterogeneity in initial conditions.

The distribution of transitory shocks, \( \psi_\gamma \), is modeled flexibly. Alongside a Gaussian benchmark case, we allow for non-Gaussian specifications that capture skewness and excess kurtosis, as well as specifications that introduce serial dependence through an MA(1) process. This flexibility enables the model to accommodate a wide range of transitory earnings innovations, including asymmetry and thick tails, which are commonly observed in administrative and survey data.

This structure nests an array of familiar models as special cases. Standard linear AR(1) models with Gaussian shocks are obtained when the conditional mean \( \mu(z_{t-1}) \) is linear, the volatility function \( \sigma(z_{t-1}) \) is constant, and the transitory components are i.i.d. draws from a normal distribution \( \psi_\gamma \). At the same time, the specification is rich enough to encompass more complex earnings processes that have been emphasized in recent empirical work. For example, \citet{ArellanoBlundellBonhomme2017} highlight the importance of nonlinear persistence and state-dependent volatility, while \citet{MeghirPistaferri2004} provide evidence of heterogeneity in both the magnitude and persistence of permanent and transitory shocks. Our framework is designed to capture precisely these features, allowing for greater flexibility while nesting a number of existing models. 

To this end, we will also consider the following extension that allows for latent time-invariant heterogeneity, 
\begin{align}
&y_t = z_t + e_t, \label{eq:mod_1_het}\\
&z_t = \mu(z_{t-1},a) + \sigma(z_{t-1},a) u_t, \label{eq:mod_2_het}\\
&(a,z_1) \sim f_{\alpha}, \quad u_t \sim \mathcal{N}(0,1), \quad e_t \,|\, a\sim \psi_\gamma(\cdot; a),\label{eq:mod_3_het}
\end{align}
where $u_t$ are serially independent, independent of $e_t$, and independent of the latent heterogeneity $a$ and initial condition $z_1$ that in turn jointly follow a distribution $f_{\alpha}$. In our application to the PSID we will estimate a particular version of this model with heterogeneity in the variance of transitory shocks.


\begin{remark}{(Conditional skewness)} Model (\ref{eq:mod_1})-(\ref{eq:mod_2})-(\ref{eq:mod_3}) imposes that $z_t$ is Gaussian given $z_{t-1}$, thus ruling out non-Gaussian features in the conditional distribution such as conditional skewness (\citealp{ArellanoBlundellBonhomme2017}). However, even with $u_t$ and $e_t$ symmetric, the model can generate conditional skewness and excess kurtosis over multiple periods ahead, even when one-period-ahead shocks are Gaussian.
\end{remark}



\section{A Review of Variational Inference}\label{sec:VI-intro}

The model given by equations (\ref{eq:mod_1})-(\ref{eq:mod_2})-(\ref{eq:mod_3}) is parametric, indexed by the parameters $\mu(\cdot)$, $\sigma(\cdot)$, $\alpha$, and $\gamma$. For conciseness we will use $\theta=(\mu(\cdot), \sigma(\cdot), \alpha)$ to denote the parameters that index the distribution of the persistent component $z_t$, whereas $\gamma$ indexes the distribution of transitory shocks $e_t$. We use the shorthand $\vartheta=(\theta,\gamma)$ to denote the vector containing all model parameters. The econometrician seeks to learn these parameters based on a sequence of outcomes $y_{1},...,y_T$, available for a collection of individuals, although we omit the individual dimension from the notation for simplicity. While we focus on the homogeneous model (\ref{eq:mod_1})-(\ref{eq:mod_2})-(\ref{eq:mod_3}) to describe variational inference, the VI approach applies naturally to model (\ref{eq:mod_1_het})-(\ref{eq:mod_2_het})-(\ref{eq:mod_3_het}) with heterogeneity; see Section \ref{sec:MA1}.

\subsection{Issues with Evaluating the Likelihood}

The log-likelihood of the observed data for a given individual can be expressed as
\begin{align}
{\cal{L}}_{\vartheta}(y_{1:T}) = \log \P_{\vartheta}(y_{1:T}) = \log \int f_\theta(z_{1:T}) \, \psi_\gamma(y_{1:T} - z_{1:T}) \, dz_{1:T},\label{eq:lik}
\end{align}
where $f_\theta(z_{1:T})$ denotes the density of $z_{1:T}=(z_1,...,z_T)$ over latent trajectories parameterized by $\theta$, and $\psi_\gamma(y_{1:T} - z_{1:T})$ is the conditional density of outcomes given $z_{1:T}$, parameterized by $\gamma$, linking latent states to observed data. Equation (\ref{eq:lik}) shows that evaluating the marginal log-likelihood ${\cal{L}}_{\vartheta}(y_{1:T})$ requires integrating over all possible latent trajectories $z_{1:T}$. However, this task is generally infeasible in realistic models of earnings dynamics.\footnote{In this paper, we view the log-likelihood function as the target objective to optimize. This presumes that the conditions for consistency of maximum likelihood are satisfied. An important necessary condition is identification, which may be challenging to establish in these models, see among others \citet{hu2008instrumental} and \citet{ArellanoBlundellBonhomme2017}.} 

To see the difficulty, consider evaluating (\ref{eq:lik}) for one individual. Since there are $T$ latent variables, $z_1,...,z_T$, numerical quadrature or grid-based methods become inaccurate as soon as there are more than a handful of periods. A common approach is to resort to simulation-based techniques such as importance sampling, particle filtering, or Markov Chain Monte Carlo, see for example \citet{creal2012survey} for a survey of particle filter methods and \citet{arellano2024heterogeneity} for an application in the context of earnings and consumption dynamics. However, while such approaches often perform well in the case of a single time series, evaluating as many integrals in (\ref{eq:lik}) as there are individuals in the sample represents a formidable challenge. As a result, state-of-the-art algorithms for nonlinear models of earnings dynamics based on these techniques are currently limited to data sets of moderate dimension, especially regarding the number of time periods.

\subsection{The Evidence Lower Bound}

To address the computational challenges of exact likelihood-based methods, we study a \textit{variational inference} (VI) approach. Instead of evaluating or maximizing the marginal log-likelihood directly, variational inference reframes the problem as an optimization of the so-called \textit{evidence lower bound} (ELBO), here again for a single individual:
\begin{equation}
 \mathcal{E}_{\vartheta,\phi}(y_{1:T}) 
= \mathbb{E}_{q_\phi(z_{1:T} \,|\, y_{1:T})}\left[\log \frac{f_\theta(z_{1:T})  \psi_\gamma(y_{1:T} - z_{1:T})}{q_\phi(z_{1:T} \,|\, y_{1:T})}\right],\label{eq:ELBO}
\end{equation}
where $q_\phi(z_{1:T} \,|\, y_{1:T})$ is some density over the latent states $z_1,...,z_T$, indexed by a parameter vector $\phi$. The density $q_{\phi}$ is called the \emph{variational posterior} density, and the expectation in (\ref{eq:ELBO}) is taken with respect to $q_\phi(z_{1:T} \,|\, y_{1:T})$ for a fixed sequence of observations $y_{1:T}$.

To understand the logic behind the maximization of (\ref{eq:ELBO}), and to see that the ELBO is indeed a \emph{lower bound} on the log-likelihood function, it is useful to introduce some notation. Interpreting $f_{\theta}$ as a \emph{prior} on the states $z_1,...,z_T$, denote the associated \emph{posterior} density as, by Bayes' rule,
\begin{equation}
p_{\vartheta}(z_{1:T}\,|\, y_{1:T})=\frac{f_{\theta}(z_{1:T})\psi_{\gamma}(y_{1:T}-z_{1:T})}{ \P_{\vartheta}(y_{1:T})}.\label{eq:true_posterior}
\end{equation}
Note that, since the likelihood function $\P_{\vartheta}(y_{1:T})$ appears in the denominator of (\ref{eq:true_posterior}), the posterior density is typically highly challenging to calculate, for the reasons mentioned in the previous subsection. 

Next, let $\mathrm{KL}[q \,\|\,  p]$ denote the Kullback–Leibler divergence between $q$ and $p$; that is,
$$\mathrm{KL}[q\,\|\,  p]=\int \log\left(\frac{q(z)}{p(z)}\right)q(z)dz.$$
Observe that the ELBO can equivalently be written as
\begin{align}
 \mathcal{E}_{\vartheta,\phi}(y_{1:T}) 
&= \mathbb{E}_{q_\phi(z_{1:T}\,|\, y_{1:T})}\left[\log \frac{f_\theta(z_{1:T})  \psi_\gamma(y_{1:T} - z_{1:T})}{q_\phi(z_{1:T}\,|\, y_{1:T})}\right]\notag\\
&=\log \P_{\vartheta}(y_{1:T}) -   \mathbb{E}_{q_\phi(z_{1:T}\,|\, y_{1:T})}\left[\log \frac{q_\phi(z_{1:T}\,|\, y_{1:T})}{\frac{f_{\theta}(z_{1:T})\psi_{\gamma}(y_{1:T}-z_{1:T})}{ \P_{\vartheta}(y_{1:T})}}\right],\notag
\end{align}
where we have used that $\P_{\vartheta}(y_{1:T})$ does not depend on the latent states $z_{1:T}$. Hence, using the expressions of the posterior density and KL divergence, we obtain the following key identity:
\begin{align}
 \mathcal{E}_{\vartheta,\phi}(y_{1:T}) 
&={\cal{L}}_{\vartheta}(y_{1:T}) -  \mathrm{KL}\big[q_\phi(z_{1:T}\,|\, y_{1:T}) \,\|\, p_{\vartheta}(z_{1:T}\,|\, y_{1:T})\big].\label{eq:lower_bound}
\end{align}

Equation (\ref{eq:lower_bound}) shows that the ELBO is equal to the log-likelihood function minus a penalty term: the KL divergence between the variational posterior density $q_{\phi}$ and the true posterior density $p_{\vartheta}$. Since the KL divergence is non-negative, (\ref{eq:lower_bound}) shows that the ELBO is indeed a lower bound on the log-likelihood. In addition, this characterization of the ELBO has two important implications. 

The first implication of (\ref{eq:lower_bound}) is that maximizing the ELBO with respect to $\phi$, for given parameters $\theta$ and $\gamma$, is equivalent to finding the variational posterior $q_{\phi}$ that is closest, in a KL sense, to the true posterior evaluated at $\vartheta=(\theta,\gamma)$. When the variational family $q_{\phi}$ is very flexible (e.g., when $\phi$ is high-dimensional), one expects the resulting KL term to be small, and the ELBO and log-likelihood to be close to each other. Indeed, in the case where the variational family includes the true posterior, ELBO (maximized with respect to $\phi$) and log-likelihood coincide. However, for a restricted variational family such as Gaussian densities, the gap between the log-likelihood and the ELBO may be substantial.


The second implication of (\ref{eq:lower_bound}) is computational. Note that the likelihood function $\P_{\vartheta}(y_{1:T})$, which is an intractable integral, does not appear in (\ref{eq:ELBO}). By replacing the log of an expectation with an expectation of logs in (\ref{eq:ELBO}), relying on the ELBO instead of the log-likelihood transforms the problem into one that is computationally tractable. Derivatives of the objective function can now be computed and averaged over, rather than requiring integration over the full latent space. This can provide important computational advantages compared to traditional methods such as the EM algorithm, as we now illustrate.

\subsection{The EM Algorithm: Alternating Optimization}

A common strategy in latent-variable models is to rely on the Expectation-Maximization (EM) algorithm. To relate the latter to the ELBO, suppose that we take $\phi=\vartheta'=(\theta',\gamma')$, for some hypothetical values of the parameters, and set $q_{\phi}$ to be the true posterior at those parameters; that is,  
\begin{equation}
q_{\phi}(z_{1:T}\,|\, y_{1:T}) = p_{\vartheta'}(z_{1:T}\,|\, y_{1:T}).\label{eq:qequalp}
\end{equation}
By (\ref{eq:lower_bound}) we have
\begin{align}
 \mathcal{E}_{\vartheta,\vartheta'}(y_{1:T})  
&={\cal{L}}_{\vartheta}(y_{1:T}) -  \mathrm{KL}\big[p_{\vartheta'}(z_{1:T}\,|\, y_{1:T}) \,\|\, p_{\vartheta}(z_{1:T}\,|\, y_{1:T})\big].\label{eq:lower_bound_EM}
\end{align}

The EM algorithm maximizes $\mathcal{E}_{\vartheta,\vartheta'}(y_{1:T}) $, by alternating between two steps:
\begin{itemize}
\item \textbf{Maximize with respect to $\vartheta'$ (E-step).}

Given $\vartheta=(\theta,\gamma)$, we see from (\ref{eq:lower_bound_EM}) that the maximum with respect to $\vartheta'$ is achieved when 
$$p_{\vartheta'}(z_{1:T}\,|\, y_{1:T})=p_{\vartheta}(z_{1:T}\,|\, y_{1:T}).$$

\item \textbf{Maximize with respect to $\vartheta$ (M-step).}

Given $\vartheta'=(\theta',\gamma')$, the bound (\ref{eq:lower_bound_EM}) attains its maximum when $\vartheta=(\theta,\gamma)$ is chosen to maximize
$$ \mathbb{E}_{p_{\vartheta'}(z_{1:T}\,|\, y_{1:T})}\left[\log f_\theta(z_{1:T})  \psi_\gamma(y_{1:T} - z_{1:T})\right].$$
\end{itemize}

Notice that the ELBO, and hence the log-likelihood, are weakly increasing in each (E,M) iteration. The EM algorithm can thus be used as an alternative to gradient-based maximization of the likelihood (\citealp{dempster1977maximum}). However, the E-step requires evaluating the exact posterior $p_{\vartheta'}(z_{1:T}\,|\, y_{1:T})$, which is generally intractable. Sampling-based approximations (using, e.g., MCMC or particle filter methods) can be employed, but they are computationally costly, difficult to differentiate, and not easily integrated into optimization routines that rely on automatic differentiation and gradient-based updates.

\subsection{Variational Inference: A Fully Differentiable Approach}

Variational inference relies on a different approach. The key idea is to replace the exact posterior in (\ref{eq:qequalp}) with a parameterized approximating family $q_{\phi}\in Q_\phi$. Instead of computing the true posterior in the E-step, we optimize the variational parameters $\phi$ to minimize the KL divergence to the (unknown) true posterior. This is equivalent to maximizing the ELBO with respect to both $\vartheta$ and $\phi$. There are two main approaches to VI in the literature, which we now review.


\paragraph{Coordinate-Ascent Variational Inference (CAVI). } 

An important input to the VI approach is the variational family. A common choice is the mean-field specification 
\begin{equation}q(z_{1:T}\,|\, y_{1:T})=\prod_{t=1}^T q_t(z_t\,|\, y_{1:T}),\label{eq:meanfield}\end{equation}
where the marginal densities $q_t$ are unrestricted. Assumption (\ref{eq:meanfield}) implies that the variational posterior factors across observations. In this case, one can show that the $q_t$'s that maximize the ELBO take an explicit form, and satisfy
\begin{equation}q_t\propto \exp(\mathbb{E}_{q_{-t}}\log p),\label{eq:CAVI}\end{equation}
where $\mathbb{E}_{q_{-t}}\log p$ is the expectation of the log-posterior density with respect to all variational marginal densities except the one in period $t$ (and $\propto$ denotes proportionality). In Coordinate-Ascent Variational Inference (CAVI), one takes advantage of (\ref{eq:CAVI}) to maximize the ELBO using coordinate ascent. By allowing for unrestricted marginal densities $q_t$, CAVI fully exploits the mean-field structure and can benefit from parallel implementation across individuals. However, the mean-field assumption can be empirically restrictive, as we will illustrate in our applications.  

\paragraph{Amortized Variational Inference with Reparameterization Trick. } 

Another common choice is to specify the variational family as a parametric density $q_{\phi}(z_{1:T}\,|\, y_{1:T})$ indexed by a parameter $\phi$ that is common across all observations in the sample. In Amortized VI, unlike in CAVI, parameters $\phi$ are shared across individuals instead of being observation-specific. A key requirement to implement the approach is that it is easy to simulate from  $q_{\phi}$, and for this reason multivariate Gaussian specifications are commonly used. At the same time, one can entertain a rich dependence on the observations $y_1,...,y_T$; for example, the mean and variance of the Gaussian variational posterior are often modeled using neural networks (e.g., \citealp{kingma2014auto}). Amortization can reduce the complexity of the variational model at the cost of losing direct parallelization, and it is well-suited to variational specifications that do not have a mean-field structure.

However, a challenge with Amortized VI is to compute gradients of the ELBO with respect to the variational parameters $\phi$, since the expectation $\mathbb{E}_{q_\phi(z_{1:T} \,|\, y_{1:T})}[\cdot]$ depends on $\phi$, preventing the use of standard gradient-based optimization. The \emph{reparameterization trick} resolves this difficulty by rewriting random draws from $q_\phi$ as deterministic transformations of the variational parameters, $\phi$, and a vector of auxiliary noise random variables, $v$,\footnote{For example, if $q_\phi(z_{1:T}\,|\, y_{1:T}) $ is a Gaussian density with mean $\mu_\phi$ and variance $\Sigma_\phi$, then $z_{1:T} = \mu_\phi + L_\phi v$ with $v \sim \mathcal{N}(0,I)$ and $L_\phi L_\phi^\top = \Sigma_\phi$, where both $\mu_{\phi}$ and $\Sigma_{\phi}$ typically depend on the observation sequence $y_{1},...,y_T$.}
\[
z_{1:T} = g(\phi, v;y_{1:T}), \quad v \sim \pi(v).
\]
This transformation allows gradients to pass through the expectation, since the gradient
\begin{align}
\nabla_\phi\mathcal{E}_{\vartheta,\phi}(y_{1:T}) 
= \mathbb{E}_{\pi(v)}\left[\nabla_\phi\left(\log \frac{f_\theta(g(\phi, v;y_{1:T}))  \psi_\gamma(y_{1:T} - g(\phi, v;y_{1:T}))}{q_\phi(g(\phi, v;y_{1:T})\,|\, y_{1:T})}\right)\right]
\end{align}
is an expectation with respect to the density $\pi(v)$, and can thus be estimated without bias using Monte Carlo draws. This makes the variational objective amenable to automatic differentiation and provides a fully differentiable alternative to sampling-based methods, allowing variational inference to be integrated with modern deep learning frameworks and deterministic or stochastic gradient optimization.

\section{Indirect Variational Inference}\label{sec:IVI}

We now show how to embed variational inference (VI) into an indirect inference approach (IVI), and contrast the theoretical properties of VI and IVI.

\subsection{A Penalized Likelihood Interpretation}

Given an i.i.d. sample drawn from $y_{1:T}$, the variational objective function can be written as
\begin{align}
 \widehat{\mathcal{E}}_{\vartheta}&=\underset{\phi}{\max}\,\widehat{\mathbb{E}}  \left[\mathcal{E}_{\vartheta,\phi}(y_{1:T})\right] \notag\\
&=\underset{\text{log-likelihood}}{\underbrace{\widehat{\mathbb{E}}  \left[{\cal{L}}_{\vartheta}(y_{1:T})\right]}} - \underset{\text{penalty}}{\underbrace{\underset{\phi}{\min}\, \widehat{\mathbb{E}}  \left[\mathrm{KL}\big[q_\phi(z_{1:T}\,|\, y_{1:T}) \,\|\, p_{\vartheta}(z_{1:T}\,|\, y_{1:T})\big]\right]}},\label{eq:QL}
\end{align}
where $\widehat{\mathbb{E}}$ indicates a sample mean of $N$ individual observations.\footnote{In (\ref{eq:QL}), the minimum over $\phi$ is outside of the sample average, thus covering amortized variational approaches.}

The variational objective function $\widehat{\mathcal{E}}_{\vartheta}$ can be interpreted as a \emph{penalized log-likelihood} function. The second term on the right-hand side of (\ref{eq:QL}) acts as a penalty, which distorts the variational inference estimates away from the maximum likelihood estimator. To see this, let $\vartheta_0$ denote the true value of the model parameter, and consider the large-sample limit 
\begin{align}
\overline{\mathcal{E}}_{\vartheta,\vartheta_0}&=\underset{\phi}{\max}\,{\mathbb{E}}_{\vartheta_0}  \left[\mathcal{E}_{\vartheta,\phi}(y_{1:T})\right].\label{eq:QL_pop}
\end{align}
While the expected log-likelihood function $\overline{\cal{L}}_{\vartheta,\vartheta_0}={\mathbb{E}}_{\vartheta_0}  \left[{\cal{L}}_{\vartheta}(y_{1:T})\right]$ is maximized at the true parameter value $\vartheta=\vartheta_0$, the population variational objective $\overline{\mathcal{E}}_{\vartheta,\vartheta_0}$ is not, and the variational estimator \begin{equation}\widehat{\vartheta}^{\rm VI}=\underset{\vartheta}{\mbox{argmax}}\, \widehat{\mathcal{E}}_{\vartheta}\label{eq:theta_VI}\end{equation}
is inconsistent for $\vartheta_0$ in large samples in general. 

The population variational objective is associated with a \emph{pseudo-true} parameter value (\citealp{westling2019beyond}), defined as
\begin{equation}\overline{\vartheta}_0=\underset{\vartheta}{\mbox{argmax}}\,\overline{{\mathcal{E}}}_{\vartheta,\vartheta_0}.\end{equation}
Under suitable regularity conditions, $\overline{\vartheta}_0$ is the large-sample limit of $\widehat{\vartheta}^{\rm VI}$. However, due to the presence of the KL penalty, $\overline{\vartheta}_0\neq \vartheta_0$ in general. This bias issue is the main challenge in applying and interpreting variational inference.

\begin{remark}{(VI bias)}
To shed further light on the source of bias of the variational estimator, decompose the large-sample limit $\overline{\mathcal{E}}_{\vartheta,\vartheta_0}$ in (\ref{eq:QL}) as
\begin{align}
\overline{\mathcal{E}}_{\vartheta,\vartheta_0}
&=\overline{\mathcal{L}}_{\vartheta,\vartheta_0} -\overline{\mathcal{KL}}_{\vartheta,\vartheta_0},\label{eq:QL_pop2}
\end{align}
and note that $\overline{\vartheta}_0$ satisfies $\nabla_{\vartheta}\overline{\mathcal{E}}_{\overline{\vartheta}_0,\vartheta_0}=0$. A local expansion around $\vartheta_0$ gives
\begin{equation}
\overline{\vartheta}_0-\vartheta_0=\left(\nabla_{\vartheta\vartheta}\overline{\mathcal{L}}_{\vartheta_0,\vartheta_0}-\nabla_{\vartheta\vartheta}\overline{\mathcal{KL}}_{\vartheta_0,\vartheta_0}\right)^{-1}\nabla_{\vartheta}\overline{\mathcal{KL}}_{\vartheta_0,\vartheta_0}+o\left(\|\overline{\vartheta}_0-\vartheta_0\|\right).
\end{equation}
This shows that, in a local approximation around $\vartheta_0$, the bias of the variational estimator depends on two terms: the derivative of the KL penalty, and the curvature of the ELBO. For a given penalty, the bias is amplified in directions where the expected ELBO is flatter, for example because identification is weaker.  
\end{remark}

\subsection{Indirect Inference}

We now introduce a strategy to de-bias the variational estimator and restore consistency. The approach is based on indirect inference (\citealp{smith1993estimating,gourieroux1993indirect}, \citealp{gallant1996moments}), and uses the variational specification as an auxiliary model.\footnote{Other approaches have been proposed as improvements to standard VI; see Appendix \ref{sec:extensions} for two of these approaches.} For exposition, we start by presenting the method at the population level, and then describe estimation based on an i.i.d. sample as well as asymptotic properties in the next subsection.

Let us define the \emph{binding function} $\vartheta\mapsto b(\vartheta)$ as  
\begin{equation}
b(\vartheta)=\underset{\widetilde\vartheta}{\mbox{argmax}}\, \overline{{\mathcal{E}}}_{\widetilde\vartheta,\vartheta},
\end{equation}
where $\overline{{\mathcal{E}}}_{\widetilde\vartheta,\vartheta}$ is the large-sample limit of the variational objective $\widehat{\mathcal{E}}_{\widetilde\vartheta}$ under ${\cal{P}}_{\vartheta}$; see (\ref{eq:QL_pop}). Hence, $b(\vartheta)$ is the probability limit of the VI estimator of $\vartheta$ under ${\cal{P}}_{\vartheta}$. By definition of the binding function, we have
\begin{equation}
b(\vartheta_0)=\overline{\vartheta}_0.\label{eq:fp}
\end{equation}
Assuming that $b(\cdot)$ is one-to-one -- a key identification assumption in indirect inference -- the IVI approach relies on the characterization of the true value $\vartheta_0$ as
\begin{equation}
\vartheta_0=\underset{\vartheta}{\mbox{argmin}}\,\left\|b(\vartheta)-\overline{\vartheta}_0\right\|^2,\label{eq:grad}
\end{equation}
where $\|\cdot\|$ is the Euclidean norm. We next describe two approaches to implement this idea.

\paragraph{Gradient descent.}

To perform gradient descent in (\ref{eq:grad}), we need the values of $b(\vartheta)$ and its derivative $\nabla_{\vartheta}b(\vartheta)$. The former can be approximated by maximizing the variational objective using data simulated under ${\cal{P}}_{\vartheta}$. As for the latter, several approaches are available, including finite difference derivatives if the dimension of $\vartheta $ is moderate. 

In Appendix \ref{app:grad} we provide more details about gradient descent. We write the expression for $\nabla_{\vartheta}b(\vartheta)$ and show it consists of two terms, one of them being typically easy to estimate using the ``reparameterization trick,'' while the second term can be more challenging to approximate when the dimension of $\phi$ is large. Given this, we also describe an alternative indirect inference approach that relies on a different, ``fixed-$q$'' binding function obtained for fixed variational estimates of $\phi$ for which gradient descent is computationally easier.


\paragraph{Fixed-point iteration.}

In the present setting, the model parameter and auxiliary parameter coincide, since both are equal to $\vartheta$, so the indirect inference problem is just-identified (see \citealp{zhang2022just}). In addition, if the variational family contains the true model posterior, then $b(\cdot)$ is the identity map since $\widehat{\vartheta}^{\rm VI}$ is consistent for $\vartheta$ under ${\cal{P}}_{\vartheta}$. It is natural to expect that, when the variational family is sufficiently rich, $b(\cdot)$ is ``not too far'' from the identity.  

These arguments motivate solving for $\vartheta_0$ in (\ref{eq:fp}) using a fixed-point iteration. Let $\kappa\in (0,1]$ be a damping factor. We implement the following fixed-point algorithm
\begin{equation} \vartheta^{(0)}=\overline{\vartheta}_0,\quad \text{and } \vartheta^{(k+1)}=\vartheta^{(k)}-\kappa \left(b(\vartheta^{(k)})-\overline{\vartheta}_0\right) \text{ for }k\geq 0,\label{eq:FP}\end{equation}
which only requires evaluating VI estimators. In practice, we recommend to perform at least a few iterations, and, if computationally feasible, to iterate until convergence subject to some numerical tolerance.

\begin{remark}{(Convergence)}
Relative to gradient descent, fixed-point iteration does not require computing derivatives, which can be computationally advantageous. However, convergence relies on a contraction mapping assumption. A sufficient condition for global convergence, irrespective of the starting value $\vartheta^{(0)}$, is that the map $\vartheta\mapsto \vartheta-\kappa b(\vartheta)$ be globally Lipschitz with constant $L<1$. If the variational family contains the true posterior then $b(\vartheta)=\vartheta$ and the condition is trivially satisfied with $L=|1-\kappa|$. When the family does not contain the true posterior, the condition is intuitively more likely to hold when the variational family is richer. In our applications we will see that fixed-point iteration based on Gaussian variational families leads to substantial improvements relative to standard variational inference.\footnote{To monitor the algorithm and assess a possible failure of the contraction assumption, one can report the quantity
$r^{(k)}=\|b(\vartheta^{(k)})-\overline{\vartheta}_0\|$
across iterations $k$, and assess whether $r^{(k)}$ is decreasing towards zero. An additional convergence check is that the ratios $r^{(k)}/r^{(k-1)}$ be persistently below $1$. When convergence appears to fail, lowering the damping factor $\kappa$ may help. There exist strategies to speed up convergence, such as Anderson acceleration (\citealp{anderson1965iterative}).}
\end{remark}

\subsection{Estimation and asymptotic properties}

So far we have presented the indirect variational inference approach at the population level. In practice, given an i.i.d. sample, $\overline{\vartheta}_0$ is estimated as $\widehat{\vartheta}^{\rm VI}$ given by (\ref{eq:theta_VI}), and the binding function $b(\cdot)$ is estimated as
$$\widehat{b}\left(\vartheta\right)=\underset{\widetilde\vartheta}{\mbox{argmax}}\, \left(\underset{\phi}{\mbox{max}}\,\widetilde{\mathbb{E}}  \left[{\mathcal{E}}_{\widetilde\vartheta,\phi}(\widetilde{y}_{1:T}(\vartheta))\right]\right) \quad \text{for all }\vartheta,$$
where $\widetilde{y}_{1:T}(\vartheta)$ are simulated draws from ${\cal{P}}_{\vartheta}$, and $\widetilde{\mathbb{E}}$ denotes a sample mean of $NM$ simulated draws, $M$ per individual observation. Hence, $\widehat{b}\left(\vartheta\right)$ is a function of simulated data, independent of the sample used to estimate $\widehat{\vartheta}^{\rm VI}$.

The indirect variational estimator of $\vartheta_0$ is then defined as
\begin{equation}
\widehat{\vartheta}^{\rm IVI}=\widehat{b}^{-1}\left(\widehat{\vartheta}^{\rm VI}\right),
\end{equation}
where
$$\widehat{b}^{-1}\left(\widehat{\vartheta}^{\rm VI}\right)=\underset{\vartheta}{\mbox{argmin}}\,\left\|\widehat b(\vartheta)-\widehat{\vartheta}^{\rm VI}\right\|^2$$
is obtained using gradient descent or fixed-point iteration. 

In Appendix \ref{app:var} we first show that, as in \citet{westling2019beyond}, the variational estimator is root-$N$ consistent and asymptotically normal for the \emph{pseudo-true value} $\overline{\vartheta}_0$,  
\begin{equation}
\sqrt{N}\left(\widehat{\vartheta}^{\rm VI}-\overline{\vartheta}_0\right)\overset{d}{\rightarrow}{\cal{N}}\left(0,V^{\rm VI}\right),
\end{equation}
where $V^{\rm VI}$ has a sandwich form and can be consistently estimated without having to compute the likelihood.

We then combine these arguments with the asymptotic analysis of indirect inference estimators in \citet{gourieroux1993indirect} to show that the indirect variational estimator is root-$N$ consistent and asymptotically normal for the \emph{true value} ${\vartheta}_0$, 
\begin{equation}
\sqrt{N}\left(\widehat{\vartheta}^{\rm IVI}-\vartheta_0\right)\overset{d}{\rightarrow}{\cal{N}}\left(0,V^{\rm IVI}\right),
\end{equation}
where $V^{\rm IVI}$ also has a sandwich form and can be estimated consistently, again avoiding the need to compute the likelihood.

\section{A Linear Gaussian Model with Closed-Form \\Likelihood and Posterior}\label{sec:benchmark}

To benchmark the variational approximations and evaluate their accuracy in a controlled setting, in this section we consider a simple linear Gaussian specification of the latent-variable model introduced in Section \ref{sec:model}. In this version, the persistent and transitory components follow linear Gaussian processes, allowing for closed-form expressions of the likelihood and posterior distribution. 
The tractability of this benchmark allows us to quantify the bias induced by different variational restrictions, implement IVI, and assess how effectively it corrects that bias.

\subsection{The Linear Gaussian Model}

We consider the following specification of the data-generating process:
\begin{align}
y_t &= z_t + e_t,\label{eq:bench_1} \\
z_t &= \rho z_{t-1} + \sigma u_t,\label{eq:bench_2} \\
z_1 &\sim \mathcal{N}(0, \sigma_{z_1}^2),  \quad u_t \sim \mathcal{N}(0,1), \quad e_t \sim \mathcal{N}(0, \sigma_e^2). \label{eq:bench_3}
\end{align}
Here, \( z_t \) is an unobserved latent state evolving as an AR(1) process with innovation variance \( \sigma^2 \), and \( y_t \) is the observed outcome (log-earnings), measured with additive Gaussian transitory shocks of variance \( \sigma_e^2 \). The initial condition is assumed to be normally distributed, \( z_1 \sim \mathcal{N}(0, \sigma_{z_1}^2) \), which allows for non-stationary initial conditions. Observations are i.i.d. across individuals (although we again omit the individual subscript from the notation for simplicity). To map this simple model to the notation of the earlier sections, here $\theta$ includes $\rho$, $\sigma$, and $\sigma_{z_1}$, and $\gamma=\sigma_e$.

\paragraph{Closed-Form Likelihood. } 
Since the model is linear and Gaussian, the joint distribution of latent states and observations is multivariate Gaussian. Specifically, we have
\begin{align}
z_{1:T} &\sim \mathcal{N}(0, \Sigma_z), \\
y_{1:T} \,|\, z_{1:T} &\sim \mathcal{N}(z_{1:T}, \sigma_e^2 I_T),
\end{align}
where $\Sigma_z$ is a function of $\rho$, $\sigma_{z_1}^2$, and $\sigma^2$ given in Appendix \ref{app:tech_details}. The marginal likelihood of the data is then obtained by integrating out the latent variables, 
\begin{align}
y_{1:T}&\sim \mathcal{N}(0, \Sigma_z + \sigma_e^2 I_T),
\end{align}
so
\begin{align}
{\cal{L}}_{\rho, \sigma, \sigma_{z_1}, \sigma_e}( y_{1:T}) = -\frac{T}{2} \log(2\pi) - \frac{1}{2} \log |\Sigma_z + \sigma_e^2 I_T| - \frac{1}{2} y_{1:T}^\top (\Sigma_z + \sigma_e^2 I_T)^{-1} y_{1:T}.\label{eq:lik_bench}
\end{align}
This expression can be evaluated exactly, as \( \Sigma_z^{-1} \) has a known tridiagonal structure that reflects the AR(1) dynamics (see Appendix \ref{app:tech_details}). 

\paragraph{True Posterior Distribution. }
Since the joint distribution of \( (z_{1:T}, y_{1:T}) \) is Gaussian, the posterior \( p(z_{1:T} \,|\, y_{1:T}) \) is also multivariate Gaussian:
\begin{equation}
z_{1:T} \,|\, y_{1:T} \sim \mathcal{N}(\mu_{z | y}, \Sigma_{z | y}),\label{eq:post_benchmark}
\end{equation}
where the mean and covariance are given by 
\begin{align}
\mu_{z | y} &= \Sigma_z (\Sigma_z + \sigma_e^2 I_T)^{-1} y_{1:T},\label{eq:muzy} \\
\Sigma_{z | y} &= \Sigma_z - \Sigma_z (\Sigma_z + \sigma_e^2 I_T)^{-1} \Sigma_z,\label{eq:sigzy}
\end{align}
with $I_T$ denoting the $T\times T$ identity matrix. These expressions show that the posterior mean is a linear smoother of the data, and that the posterior covariance shrinks relative to the prior covariance as a function of the signal-to-noise ratio. Note that, while the posterior mean $\mu_{z | y}$ is a linear function of the observed data $y_{1:T}$, the posterior covariance $\Sigma_{z | y}$ depends only on the model parameters and not on the observations sequence. 

The posterior covariance matrix $\Sigma_{z | y}$ inherits the Markovian structure of the latent process. Its inverse, the posterior precision matrix \( \Omega^{\rm post} = \Sigma_{z | y}^{-1} \), is obtained as the sum of the prior precision $ \Sigma_z^{-1}$ (which is tridiagonal due to the AR(1) dynamics) and the observation precision $ \sigma_e^{-2} I_T$:
\begin{align}
\Omega^{\rm post} = \Sigma_z^{-1} + \sigma_e^{-2} I_T.\label{eq:omegapost}
\end{align}
The posterior precision matrix remains sparse and tridiagonal, reflecting the fact that each state $z_t$ interacts directly only with its neighbors $z_{t-1}$ and $z_{t+1}$ in the likelihood.

\subsection{Variational Posteriors in the Linear Gaussian Model}

Implementing variational inference requires defining the variational posterior family $q_{\phi}$ to optimize over. We describe here the variational posteriors that we will use in the benchmark model and in the nonlinear frameworks in subsequent sections.

\paragraph{Gaussian Variational Posterior. } 
Since by (\ref{eq:post_benchmark}) the true posterior distribution \( p(z_{1:T} \,|\, y_{1:T}) \) in the linear-Gaussian model is itself multivariate normal, the Gaussian family is a natural choice for $q_{\phi}\in{\mathcal{Q}}_{\phi}$. That is, for \( \mu_q \in \mathbb{R}^T \) and \( \Sigma_q \in \mathbb{R}^{T \times T} \) we set the variational parameter as $\phi=(\mu_q,\Sigma_q)$, and the variational posterior density $q_{\phi}$ as the $\mathcal{N}(\mu_q, \Sigma_q)$ density: 
\begin{equation}q_{\phi}(z_{1:T}\,|\, y_{1:T})=\frac{1}{(2\pi)^{T/2}\,|\Sigma_q|^{1/2}}
\exp\!\left(
-\tfrac{1}{2}(z_{1:T}-\mu_q)^\top \Sigma_q^{-1}(z_{1:T}-\mu_q)
\right).\label{eq:q_gauss}\end{equation}
In the linear Gaussian model, the true posterior density belongs to the Gaussian family ${\mathcal{Q}}_{\phi}$, which thus satisfies the ``truth-in-class'' property -- a property that will fail in the nonlinear models we will study in subsequent sections.

In the linear-Gaussian model, with the choice (\ref{eq:q_gauss}) of variational density, the evidence lower bound (ELBO) is available in closed form, as
\begin{align}
{\mathcal{E}}_{\rho,\sigma,\sigma_{z_1},\sigma_e,\mu_q,\Sigma_q}(y_{1:T}) &= -\frac{1}{2} \bigg[ \sigma_e^{-2}\text{tr}(\Sigma_q + (\mu_q {-} y_{1:T})(\mu_q - y_{1:T})^\top) + \text{tr}(\Sigma_z^{-1} (\Sigma_q + \mu_q \mu_q^\top)) \notag\\
&\quad\quad\quad+\log |\Sigma_z|+T\log \sigma_e^2- \log |\Sigma_q| \bigg] + \text{constant},\label{eq:elbo_benchmark}
\end{align}
where the constant term is irrelevant for optimization. The objective function in (\ref{eq:elbo_benchmark}) is differentiable and can be optimized using standard gradient-based methods. 

For implementation, we parameterize the precision matrix \( \Sigma_q^{-1} \) directly using a Cholesky factor, and we compute gradients using the reparameterization trick. Furthermore, we rely on amortized variational inference to specify how both $\mu_q$ and $\Sigma_q$ depend on observations $y_{1:T}$. Following a common practice in the literature (e.g., \citealp{kingma2014auto}), we specify the elements of $\mu_q(y_{1:T})$ and the Cholesky factor of $\Sigma_q(y_{1:T})$ as feedforward neural networks. In particular, this implies that the model used in estimation features a very large number of parameters. We will rely on the same approach for nonlinear models in subsequent sections. Appendix \ref{app:comp} provides further details about implementation.

\paragraph{Restricted Variational Posteriors. }
The Gaussian variational family in (\ref{eq:q_gauss}) leaves the parameters $\mu_q$ and $\Sigma_q$ almost unrestricted.\footnote{Except for some mild restrictions on their dependence on $y_1,...,y_T$, modeled using flexible neural networks.} However, it may be appealing to impose some of the features of the model's true posterior on the variational family. By reducing the number of parameters involved, restrictions on the variational family may improve estimation accuracy.

We consider three types of restrictions on the Gaussian variational family: under \emph{tridiagonal precision}, a \emph{hidden Markov} structure, and \emph{diagonal precision}, respectively. To motivate the first restriction, note that the true posterior in the linear-Gaussian AR(1) benchmark model has a \emph{tridiagonal} precision matrix. It may thus be appealing to impose this restriction on $\Sigma_q^{-1}$. 

To motivate the second restriction, note that the model has a \emph{hidden Markov} structure, so the true posterior density satisfies:
\[
p(z_{1:T} \,|\, y_{1:T}) = p(z_1 \,|\, y_{1:T}) \prod_{t=2}^{T} p(z_t \,|\, z_{t-1}, y_{t:T}),
\]
where the last part only depends on future observations $y_t,...,y_T$. Indeed, conditional on \( z_{t-1} \), past outcomes \( y_{1:t-1} \) are redundant in predicting \( z_t \), since their information is mediated entirely through \( z_{t-1} \). This restriction can be incorporated into the variational family as well, 
\begin{align}
q(z_{1:T} \,|\, y_{1:T}) = q(z_1 \,|\, y_{1:T}) \prod_{t=2}^T q(z_t \,|\, z_{t-1}, y_{t:T}).\label{eq:var_HMM}
\end{align}
Modeling $q(z_t \,|\, z_{t-1}, y_{t:T})$ as conditionally Gaussian delivers a variational posterior density for $z_{1:T}$ that is not Gaussian but encodes the hidden Markov structure present in the model.

We will also consider a third type of restriction, imposing that the variational posterior covariance, and hence the posterior precision as well, are \emph{diagonal}. This restriction imposes that all off-diagonal elements of $\Sigma_q^{-1}$ are equal to zero, hence requiring that, under  $q_{\phi}$, the $z_t$'s are independent of each other. While easy to enforce in practice, this restriction \emph{does not hold} under the model, since the first off-diagonal elements of $\Omega^{\rm post}$ are all non-zero except when $\rho=0$. Independence assumptions (so-called ``mean-field'' approximations) are often imposed in applications of variational inference. However, we will see that in models of earnings dynamics they can lead to substantial biases on parameter estimates.

\subsection{Results for the Linear Gaussian Model }

To evaluate the performance of different variational posterior approximations, we conduct simulation experiments using data generated from model (\ref{eq:bench_1})-(\ref{eq:bench_2})-(\ref{eq:bench_3}). This allows us to directly assess how well different estimators recover the true parameters. 

\paragraph{Data-Generating Process. } 
In the benchmark design, the latent process $z_t$ follows an AR(1) with persistence parameter $\rho=0.9$. The initial state has standard deviation $\sigma_{z_1}=0.39$, innovations are scaled by $\sigma=0.2$, and the transitory component is normally distributed with standard deviation $\sigma_e=0.23$. These parameters are chosen to approximately match estimates based on PSID data (e.g., \citealp{BlundellEtAl2008}, \citealp{ArellanoBlundellBonhomme2017}). We simulate a panel with $N = 30{,}000$ individuals over $T = 6$ periods.

\paragraph{Estimation Model and Variational Family. } 
While the true data-generating process is governed by a first-order linear Gaussian model, in the estimation procedure we allow for greater flexibility by specifying the conditional mean and volatility using second-order polynomials:
\[
\mu(z_{t-1}) = \mu_0 + \mu_1 z_{t-1} + \mu_2 z_{t-1}^2, \quad
\sigma(z_{t-1}) = \log\left(1+ \exp\left( \sigma_0 + \sigma_1 z_{t-1} + \sigma_2 z_{t-1}^2 \right)\right).
\]
We also specify the distribution of the transitory component \( e_t \) and the initial state \( z_1 \) to be normal. This setup allows us to assess whether the variational approach can correctly recover the linear DGP by estimating the other polynomial coefficients \( \mu_0, \mu_2 \) and \( \sigma_1, \sigma_2 \) to be close to zero. Lastly, we compare the results for various choices of variational posterior families, as described above.

\paragraph{Results. } 

Our findings are presented in Table~\ref{ar1n_parameter_table}. We find that variational inference based on an unrestricted Gaussian variational posterior performs well in recovering the parameters of the model (first row in Table~\ref{ar1n_parameter_table}). Imposing a tridiagonal precision matrix leads to very similar estimates (second row). Imposing a hidden Markov structure on the variational posterior, as in (\ref{eq:var_HMM}), shows equivalent performance to the unrestricted and tridiagonal specifications (third row). Hence, imposing restrictions on the variational posterior that are satisfied by the true posterior leads to accurate inference in this setting.

\begin{table}[!t]
\centering
\caption{Parameter Estimates in the Linear Gaussian Model}
\label{ar1n_parameter_table}
\setlength{\tabcolsep}{5pt}%
\resizebox{\ifdim\width>\linewidth\linewidth\else\width\fi}{!}{%
\begin{tabular}{lcccccccc}
\toprule
 & \multicolumn{3}{c}{$\mu(z_{t-1})$} & \multicolumn{3}{c}{$\sigma(z_{t-1})$} & $f_\alpha(z_1)$ & $\psi_\gamma(e_t)$ \\
\cmidrule(l{2pt}r{2pt}){2-4}\cmidrule(l{2pt}r{2pt}){5-7}\cmidrule(l{2pt}r{2pt}){8-8}\cmidrule(l{2pt}r{2pt}){9-9}
Parameter & $\mu_0$ & $\mu_1$ & $\mu_2$ & $\sigma_0$ & $\sigma_1$ & $\sigma_2$ & $\sigma_{z_1}$ & $\sigma_e$ \\
\midrule
DGP & $\phantom{-}0.00$ & $\phantom{-}0.90$ & $\phantom{-}0.00$ & $-1.48$ & $\phantom{-}0.00$ & $\phantom{-}0.00$ & $\phantom{-}0.39$ & $\phantom{-}0.23$ \\
\addlinespace
\multicolumn{9}{l}{\textit{Variational posterior}} \\
(1) unrestr.\ Gaussian & $\phantom{-}0.00$ & $\phantom{-}0.89$ & $\phantom{-}0.00$ & $-1.45$ & $-0.01$ & $-0.01$ & $\phantom{-}0.39$ & $\phantom{-}0.23$ \\
(2) tridiagonal & $\phantom{-}0.00$ & $\phantom{-}0.90$ & $\phantom{-}0.00$ & $-1.46$ & $\phantom{-}0.01$ & $\phantom{-}0.00$ & $\phantom{-}0.39$ & $\phantom{-}0.23$ \\
(3) hidden Markov & $\phantom{-}0.00$ & $\phantom{-}0.90$ & $-0.01$ & $-1.46$ & $\phantom{-}0.02$ & $-0.01$ & $\phantom{-}0.39$ & $\phantom{-}0.23$ \\
(4) diagonal & $\phantom{-}0.00$ & $\phantom{-}0.76$ & $\phantom{-}0.00$ & $-0.99$ & $\phantom{-}0.00$ & $\phantom{-}0.00$ & $\phantom{-}0.44$ & $\phantom{-}0.13$ \\
\addlinespace
\multicolumn{9}{l}{\textit{Indirect Variational Inference}} \\
(5) diagonal (IVI) & $\phantom{-}0.00$ & $\phantom{-}0.90$ & $\phantom{-}0.00$ & $-1.47$ & $\phantom{-}0.00$ & $\phantom{-}0.00$ & $\phantom{-}0.39$ & $\phantom{-}0.23$ \\
\addlinespace
\multicolumn{9}{l}{\textit{Ignoring transitory shocks}} \\
(6) & $\phantom{-}0.00$ & $\phantom{-}0.69$ & $\phantom{-}0.00$ & $-0.84$ & $\phantom{-}0.00$ & $\phantom{-}0.01$ & $\phantom{-}0.46$ & -- \\
\bottomrule
\end{tabular}%
}
\\ \vspace{0.2cm} \figurenote{The table reports the true parameter values of the simulated DGP alongside the estimates obtained under four variational-posterior specifications and under Indirect Variational Inference (IVI) applied to the diagonal posterior. The estimation model allows for quadratic terms in both the conditional mean $\mu(z_{t-1})$ and the conditional volatility $\sigma(z_{t-1})$. The final specification abstracts from transitory shocks. Linear Gaussian DGP, $N = 30{,}000$, $T = 6$.}
\end{table}


However, imposing that $\Sigma_q$ is diagonal, which is a feature that is \emph{not} present in the true posterior under the DGP, leads to biases (fourth row in Table~\ref{ar1n_parameter_table}). In particular, the estimate of persistence of $z_t$ is attenuated, with the autoregressive coefficient estimated at 0.76, compared to the true value of 0.9 in the DGP. This downward bias reflects the inability of the diagonal specification to capture temporal dependencies in the latent states. In addition, the transitory shock standard deviation is underestimated, while the innovation standard deviation of the latent process is overestimated. 

To correct for the biases implied by the mean-field specification, we implement the indirect variational inference method (fifth row in Table~\ref{ar1n_parameter_table}). We use the fixed-point iteration (\ref{eq:FP}), for $\kappa=0.6$, starting from the diagonal (mean-field) parameter estimates and iterating 10 times. The IVI method fully corrects the parameter biases in this case, underscoring the usefulness of relying on variational specifications as auxiliary models. Figure~\ref{ar1n_trajectory_fixed_point} shows the path across fixed-point iterations for the parameters $\rho$ and $\sigma_{e}$. The arrows show the update direction $\widehat{\vartheta}^{\rm VI}-\widehat b(\vartheta)$, while the teal line traces the realized IVI trajectory starting from the mean-field VI estimate (orange diamond). The trajectory moves almost directly toward the true parameter value (blue triangle). After only a few iterations, the IVI correction brings the estimate close to the truth.

\input{figures/ar1n_trajectory_fixed_point}

Lastly, to put these findings into perspective, we report in the last row of Table~\ref{ar1n_parameter_table} estimates that are based on a Gaussian model without a transitory component. Even in the benchmark AR(1) model, omitting transitory components in observed log earnings leads to serious estimation bias. Specifically, the estimated persistence parameter is attenuated (as expected in the presence of classical measurement error) and the innovation volatility is systematically overestimated. This underscores the critical importance of accounting for transitory shocks in models of earnings dynamics, as has been extensively demonstrated in the literature.

\subsection{Further Insights on the Mean-Field Approximation}

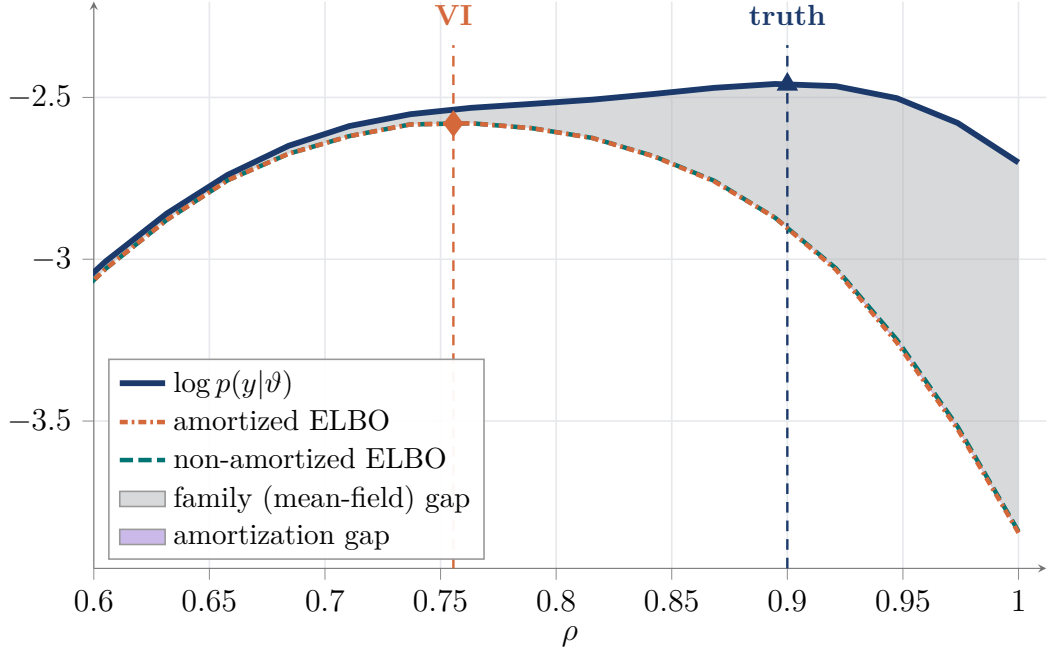
\begin{figure}[!t]
\centering
\begin{tikzpicture}
\begin{axis}[
    scale only axis,
    width=12.6cm, height=7.5cm,
    axis lines=left,
    axis line style={draw=black!55},
    title={},
    title style={font=\normalsize, yshift=2pt},
    label style={font=\normalsize},
    tick label style={font=\normalsize},
    xlabel={$\rho$},
    xlabel style={yshift=2pt},
    xmin=0.6000, xmax=1.0120,
    ymin=-3.9552, ymax=-2.2042,
    xtick distance=0.05,
    ytick distance=0.5,
    tick align=outside, tick style={draw=black!45},
    grid=both,
    major grid style={draw=cgrid, line width=0.6pt},
    minor tick num=0,
    clip=true,
    legend cell align=left,
    legend style={
        at={(0.015,0.015)}, anchor=south west,
        draw=black!40, fill=white, fill opacity=0.95, text opacity=1,
        line width=0.6pt, font=\small, row sep=1pt,
    },
]

\addplot[name path=plogp, draw=none, forget plot] coordinates {(1.000000,-2.701135) (0.973684,-2.578982) (0.947368,-2.502367) (0.921053,-2.465243) (0.894737,-2.458416) (0.868421,-2.470362) (0.842105,-2.489410) (0.815789,-2.506900) (0.789474,-2.519954) (0.763158,-2.532214) (0.736842,-2.551935) (0.710526,-2.588562) (0.684211,-2.649669) (0.657895,-2.739456) (0.631579,-2.858776) (0.605263,-3.005991) (0.578947,-3.178000) (0.552632,-3.371084) (0.526316,-3.581461) (0.500000,-3.805595)};
\addplot[name path=pnon,  draw=none, forget plot] coordinates {(1.000000,-3.841062) (0.973684,-3.517441) (0.947368,-3.248283) (0.921053,-3.028202) (0.894737,-2.871962) (0.868421,-2.757660) (0.842105,-2.681143) (0.815789,-2.625469) (0.789474,-2.595033) (0.763158,-2.579863) (0.736842,-2.583515) (0.710526,-2.619690) (0.684211,-2.674145) (0.657895,-2.756135) (0.631579,-2.878486) (0.605263,-3.028673) (0.578947,-3.208500) (0.552632,-3.397372) (0.526316,-3.605825) (0.500000,-3.843478)};
\addplot[name path=pamo,  draw=none, forget plot] coordinates {(1.000000,-3.845315) (0.973684,-3.525221) (0.947368,-3.257627) (0.921053,-3.032757) (0.894737,-2.871962) (0.868421,-2.759723) (0.842105,-2.681821) (0.815789,-2.625645) (0.789474,-2.595033) (0.763158,-2.579863) (0.736842,-2.583515) (0.710526,-2.619691) (0.684211,-2.674145) (0.657895,-2.756135) (0.631579,-2.878486) (0.605263,-3.028673) (0.578947,-3.210235) (0.552632,-3.397372) (0.526316,-3.605825) (0.500000,-3.870493)};
\addplot[cfam, opacity=0.5, forget plot] fill between[of=plogp and pnon];
\addplot[camg, opacity=0.85, forget plot] fill between[of=pnon and pamo];

\draw[clogp, dash pattern=on 4pt off 3pt, line width=0.9pt]
    (axis cs:0.900000,-3.9552) -- (axis cs:0.900000,-2.3384);
\draw[camo, dash pattern=on 4pt off 3pt, line width=0.9pt]
    (axis cs:0.755547,-3.9552) -- (axis cs:0.755547,-2.3384);
\node[clogp, font=\small\bfseries, anchor=south]
    at (axis cs:0.900000,-2.3101) {truth};
\node[camo, font=\small\bfseries, anchor=south]
    at (axis cs:0.755547,-2.3101) {VI};

\addplot[clogp, line width=2.2pt, solid, forget plot]
    coordinates {(1.000000,-2.701135) (0.973684,-2.578982) (0.947368,-2.502367) (0.921053,-2.465243) (0.894737,-2.458416) (0.868421,-2.470362) (0.842105,-2.489410) (0.815789,-2.506900) (0.789474,-2.519954) (0.763158,-2.532214) (0.736842,-2.551935) (0.710526,-2.588562) (0.684211,-2.649669) (0.657895,-2.739456) (0.631579,-2.858776) (0.605263,-3.005991) (0.578947,-3.178000) (0.552632,-3.371084) (0.526316,-3.581461) (0.500000,-3.805595)};
\addplot[cnon, line width=1.7pt, dash pattern=on 5pt off 2pt, forget plot]
    coordinates {(1.000000,-3.841062) (0.973684,-3.517441) (0.947368,-3.248283) (0.921053,-3.028202) (0.894737,-2.871962) (0.868421,-2.757660) (0.842105,-2.681143) (0.815789,-2.625469) (0.789474,-2.595033) (0.763158,-2.579863) (0.736842,-2.583515) (0.710526,-2.619690) (0.684211,-2.674145) (0.657895,-2.756135) (0.631579,-2.878486) (0.605263,-3.028673) (0.578947,-3.208500) (0.552632,-3.397372) (0.526316,-3.605825) (0.500000,-3.843478)};
\addplot[camo, line width=1.7pt, dash pattern=on 3pt off 1.5pt on 1pt off 1.5pt, forget plot]
    coordinates {(1.000000,-3.845315) (0.973684,-3.525221) (0.947368,-3.257627) (0.921053,-3.032757) (0.894737,-2.871962) (0.868421,-2.759723) (0.842105,-2.681821) (0.815789,-2.625645) (0.789474,-2.595033) (0.763158,-2.579863) (0.736842,-2.583515) (0.710526,-2.619691) (0.684211,-2.674145) (0.657895,-2.756135) (0.631579,-2.878486) (0.605263,-3.028673) (0.578947,-3.210235) (0.552632,-3.397372) (0.526316,-3.605825) (0.500000,-3.870493)};

\addplot[only marks, mark=triangle*, mark size=4.0pt,
    mark options={fill=clogp, draw=clogp, line width=0.7pt}, forget plot]
    coordinates {(0.900000,-2.459781)};
\addplot[only marks, mark=diamond*, mark size=4.5pt,
    mark options={fill=camo, draw=camo, line width=0.7pt}, forget plot]
    coordinates {(0.755547,-2.580919)};

\addlegendimage{clogp, line width=2.2pt, solid}
\addlegendentry{$\log p(y|\vartheta)$}
\addlegendimage{camo, line width=1.7pt, dash pattern=on 3pt off 1.5pt on 1pt off 1.5pt}
\addlegendentry{amortized ELBO}
\addlegendimage{cnon, line width=1.7pt, dash pattern=on 5pt off 2pt}
\addlegendentry{non-amortized ELBO}
\addlegendimage{area legend, fill=cfam, fill opacity=0.5, draw=none}
\addlegendentry{family (mean-field) gap}
\addlegendimage{area legend, fill=camg, fill opacity=0.85, draw=none}
\addlegendentry{amortization gap}

\end{axis}
\end{tikzpicture}
\caption{Log-Likelihood, Amortization Gap and Family (Mean-Field) Gap}
\label{ar1n_likelihood_elbo}
\vspace{0.2cm} \figurenote{Log-likelihood (solid blue), amortized (orange dash-dotted) and non-amortized (teal dashed) mean-field ELBO along the line $\vartheta(\iota)=\vartheta_0+\iota(\widehat\vartheta^{\rm VI}-\vartheta_0)$ from the DGP truth to the VI optimum, plotted against $\rho$. The log-likelihood peaks near the truth $\rho_0=0.90$, the amortized ELBO near the VI pseudo-true value $\overline{\rho}_0=0.76$. Purple: amortization gap (non-amortized minus amortized ELBO); gray: family gap (log-likelihood minus non-amortized ELBO). Linear Gaussian DGP, $N = 30{,}000$, $T = 6$.}
\end{figure}

Table~\ref{ar1n_parameter_table} shows that variational inference based on a diagonal, mean-field specification fails at recovering the true parameters. To shed further light on this finding, in Figure~\ref{ar1n_likelihood_elbo} we plot the log-likelihood function (solid blue) and the ELBO (dash-dotted orange), plotted against the value of $\rho$. While the log-likelihood is maximized at the true value $\rho_0=0.90$ (in the population problem), the ELBO is maximized at $\overline\rho_0=0.76$. In this model we also compare our amortized solution to the CAVI solution obtained by letting the variational parameters be observation-specific; the difference between the two is the ``amortization gap.''

Figure~\ref{ar1n_likelihood_elbo} shows that the bias reflects a systematic distortion of the variational objective rather than numerical optimization error. The bias is also not primarily driven by amortization since the non-amortized optimal mean-field ELBO lies only slightly above the amortized ELBO, revealing a negligible amortization gap. Instead, the dominant source of the discrepancy is the family, or mean-field, gap. The resulting bias therefore reflects approximation error from a posterior family that is too restrictive.

\begin{figure}[!t]
\centering
\begin{tikzpicture}
\begin{axis}[
    width=8cm, height=5cm,
    scale only axis=true,
    xmin=0.54, xmax=1.0,
    ymin=0.36, ymax=1.0,
    xtick distance=0.1,
    ytick distance=0.1,
    xlabel={$\rho_0$},
    ylabel={$\hat\rho$},
    label style={font=\normalsize},
    tick label style={font=\normalsize},
    axis lines=left,
    enlargelimits=false,
    grid=major,
    grid style={gray!20},
    tick align=outside,
]

\addplot[domain=0.54:1.0, samples=2, dotted, gray, thick] {x}
    node[sloped, pos=0.93, fill=white, inner sep=1.5pt, font=\small, text=black] {$45^\circ$};

\addplot[domain=0.54:1.0, samples=2, camo, thick, dashed] {0.7555468082427979}
    node[pos=0.2, above, text=camo, font=\normalsize\bfseries] {VI (0.76)};

\addplot[blue!55!black, thick, dashed] coordinates {(0.9,0.36) (0.9,1.0)}
    node[pos=0.07, fill=white, inner sep=2pt, text=blue!55!black, font=\normalsize\bfseries] {truth (0.90)};

\addplot[
    teal!90!black, thick, mark=*, mark size=1.6pt,
] coordinates {
    (0.55,0.3736295700073242)
    (0.6000000000000001,0.4203275740146637)
    (0.65,0.4669641852378845)
    (0.7000000000000001,0.5187948942184448)
    (0.75,0.5736395120620728)
    (0.8,0.6324478387832642)
    (0.8500000000000001,0.6933422088623047)
    (0.9,0.7557213306427002)
    (0.95,0.8201918601989746)
    (1.0,0.8845683336257935)
};

\end{axis}
\end{tikzpicture}
\caption{Slice of the Binding Function for $\rho$}
\label{ar1n_binding_function}
\vspace{0.2cm} \figurenote{Binding function $\widehat b_\rho(\vartheta)$ for the AR(1) coefficient (solid teal): the mean-field VI estimate of $\rho$ on data $\widetilde y_{1:T}(\vartheta)$ simulated under $\mathcal P_{\vartheta}$. Along the horizontal axis only the true $\rho_0$ varies, while the other coordinates of $\vartheta$ stay at their true values. At $\rho_0=0.90$ (vertical dashed blue) the binding function returns $\widehat\rho^{\rm VI}=0.76$ (horizontal dashed orange). Linear Gaussian DGP, $N = 30{,}000$, $T = 6$.}
\end{figure}
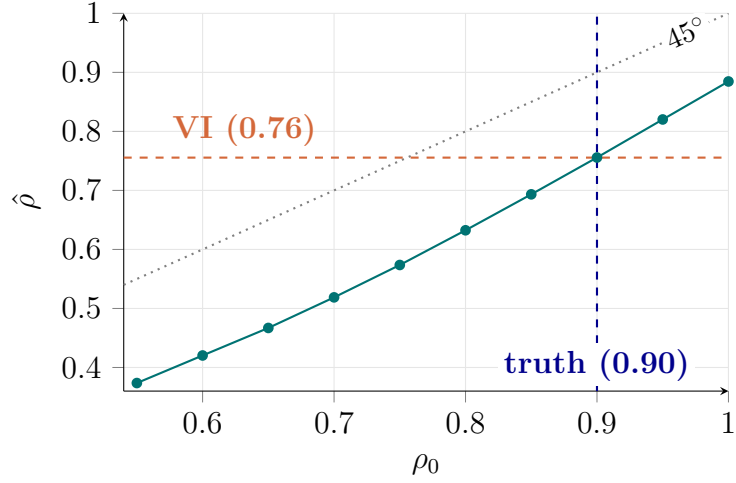

To complement the bias illustration in Figure~\ref{ar1n_likelihood_elbo}, it is useful to report some analytical calculations. Consider a Gaussian variational density with parameters $\phi=(\mu_q,\Sigma_q)$, where $\mu_q$ is unrestricted and, following  the mean-field approach, $\Sigma_q$ is diagonal.\footnote{It is well-known that the Gaussian variational family is not restrictive in this case. Suppose that (\ref{eq:meanfield}) holds, and that one maximizes the ELBO with respect to unrestricted marginal densities $q_1,...,q_T$ without imposing they are Gaussian. In that case, by (\ref{eq:CAVI}), and since the log-posterior is quadratic in $z_1,...,z_T$ by (\ref{eq:post_benchmark}), it follows that $q_t$ is Gaussian (see \citealp{bishop2006pattern}, Chapter 10).} Then, we show in Appendix \ref{app:tech_details} that the penalized log-likelihood in (\ref{eq:QL}) equals
\begin{align}
&\ {\mathcal{E}}^*_{\rho, \sigma, \sigma_e, \sigma_{z_1}}(y_{1:T}) \notag\\&= \underset{\text{log-likelihood}}{\underbrace{-\frac{T}{2} \log(2\pi) - \frac{1}{2} \log |\Sigma_z + \sigma_e^2 I_T| - \frac{1}{2} y_{1:T}^\top (\Sigma_z + \sigma_e^2 I_T)^{-1} y_{1:T}}}-\underset{\text{penalty}}{\underbrace{\frac{1}{2}\log \frac{|\mbox{diag}\, \Omega^{\rm post}|}{|\Omega^{\rm post}|}}},\label{eq:QL_bench}
\end{align}
where $\mbox{diag}\, \Omega^{\rm post}$ is the diagonal of $\Omega^{\rm post}$ given in (\ref{eq:omegapost}).

The penalty term in (\ref{eq:QL_bench}) is always non-negative and depends only on the parameters, not on the data. It captures dependence in $\Omega^{\rm post}$, and it is equal to zero when $\Omega^{\rm post}$ is diagonal, that is, when the latent states are independent in the true posterior. In the linear Gaussian model, the penalty is an increasing function of the autoregressive coefficient $|\rho|$, consistent with Figure~\ref{ar1n_likelihood_elbo}. When persistence is high, as is typically observed in earnings data, mean-field approximations based on a diagonal variational posterior covariance matrix distort the variational estimator away from the maximum likelihood estimator. 

Lastly, it is important to note that a sizable VI bias, as illustrated in Figure~\ref{ar1n_likelihood_elbo}, is not necessarily an obstacle to reliable parameter estimation, provided that one uses the variational specification as an auxiliary model. To see this, in Figure~\ref{ar1n_binding_function} we show the binding function (as a function of $\rho_0$), which gives the value of the variational estimate for each true parameter value. While the binding function, in teal, lies consistently below the 45-degree line -- thus indicating bias -- it is one-to-one, and one can \emph{invert} it by finding the parameter value on the x-axis that delivers the VI point-estimate in the data, corresponding to the point where the binding function and the orange horizontal line intersect. The solution is $0.90$ -- virtually identical to the true $\rho_0$ in this case. What matters to the success of IVI is whether the binding function is sufficiently regular to be inverted.

\section{A Nonlinear Non-Gaussian Model }\label{sec:nonlinear-model}

We now extend the linear Gaussian model in two directions. First, we introduce nonlinearity in the latent process by incorporating a nonlinear conditional mean and state-dependent volatility. Second, we relax the Gaussian assumptions for the initial latent state and the transitory component.

\subsection{The Nonlinear Model}

We focus on model (\ref{eq:mod_1})-(\ref{eq:mod_2})-(\ref{eq:mod_3}), where we allow both the latent mean function $\mu(z_{t-1})$ and the state-dependent volatility function $\sigma(z_{t-1})$ to be nonlinear, while adding non-Gaussian features in both the initial state $z_1$ and the transitory shock $e_t$. The nonlinear functions $\mu(z_{t-1})$ and $\sigma(z_{t-1})$ capture deviations from the linear AR(1) process that we studied in the previous section. 

To flexibly depart from Gaussianity, we model the transitory shock $e_t$ and the initial latent state $z_1$ as \emph{sinh–arcsinh} transformations of a standard normal random variable  \citep{Jones2014}. The resulting cumulative distribution functions are
\begin{align}
   \Psi_\gamma(e) &= \Phi\!\left( \sinh\!\left( \gamma_2\,\operatorname{asinh}\left(\tfrac{e-\gamma_0}{\gamma_1}\right) -\gamma_3 \right)\right), \label{eq:sinh-e}\\
   F_{\alpha}(z_1) &= \Phi\!\left( \sinh\!\left( \alpha_2\,\operatorname{asinh}\left(\tfrac{z_1-\alpha_0}{\alpha_1}\right)  -\alpha_3 \right)\right), \label{eq:sinh-z1}
\end{align}
where $(\gamma_0,\gamma_1, \gamma_2,\gamma_3)$ and $(\alpha_0,\alpha_1, \alpha_2,\alpha_3)$ are location, scale, tail-shape, and skewness parameters for $e_t$ and $z_1$, respectively, and $\Phi(\cdot)$ denotes the standard normal CDF. Values $\gamma_2<1$ (or $\alpha_2<1$) generate heavier tails and positive excess kurtosis relative to the Gaussian benchmark, while values above one yield thinner tails. The scale parameters $\gamma_1$ and $\alpha_1$ control dispersion, allowing standard deviation and tail behavior to vary independently. The parameters $\gamma_3$ and $\alpha_3$ control skewness, while $\gamma_0$ and $\alpha_0$ govern location. This parameterization provides a tractable way to match empirical distributions with skewness and excess kurtosis.

The resulting model departs from the benchmark state-space setting in several important ways: the presence of nonlinear mean and volatility breaks the linear-Gaussian assumption, the non-Gaussianity in \( z_1 \) and \( e_t \) implies that the posterior distribution over the latent path \( z_{1:T} \) is no longer Gaussian, and its precision matrix is no longer tridiagonal in general. Moreover, the stochastic volatility term \( \sigma(z_{t-1}) \) introduces multiplicative heteroskedasticity, which creates nonlinearities in the conditional likelihood and undermines the conjugacy that made posterior calculations analytically tractable in the linear Gaussian case. In this context, our goal is to assess whether a variational inference approach based on a Gaussian variational family -- with or without additional structure -- is able to correctly recover the model parameters despite the non-Gaussianity of the posterior.

\paragraph{Parameterization. }
To calibrate the DGP, we choose parameters for the transitory component $e_t$ to match a standard deviation of $0.16$ and a kurtosis of $10$, and for the initial latent state $z_1$ to match a standard deviation of $0.4$, a skewness close to $0$, and a kurtosis of $3.3$, in line with \citet{ArellanoBlundellBonhomme2017}. This yields parameter values $(\gamma_0,\gamma_1, \gamma_2,\gamma_3) = (0,0.033, 0.47,0)$ for $e_t$ and $(\alpha_0,\alpha_1, \alpha_2,\alpha_3) = (0,0.34, 0.89,0)$ for $z_1$. For the nonlinear conditional mean function $\mu(z_{t-1})$, we adopt a specification designed to capture a \emph{hockey stick} shape: the left tail of the distribution is nearly flat (mimicking a floor in earnings changes, e.g., due to minimum wages), while the right side increases with a persistence parameter close to $0.9$. Specifically, in the DGP we assume that
\begin{equation}\label{eq:nonlinear_mu}
\mu(z_{t-1}) \;=\; -0.25 + 0.1 \log \left[ 1+ \exp\left(\tfrac{1}{0.1} \left(0.9 z_{t-1} + 0.25 \right)\right) \right],
\end{equation}
which is strictly increasing and smooth. We maintain a quadratic innovation volatility function of the form
\begin{equation}\label{eq:nonlinear_sigma}
  \sigma(z_{t-1}) =  \log\left(1+ \exp\left( -1.8 + 0.35 z_{t-1}^2 \right)\right).
\end{equation}
Finally, we simulate $N=30{,}000$ individuals over $T=6$ periods. The parameters of the model are summarized in Table~\ref{hockeystick_parameter_table}.

In estimation, we allow for more flexible functional forms:
\begin{align}
\mu(z_{t-1}) &= \eta_0 + \eta_1 \log \left[ 1+ \exp\left(\tfrac{1}{\alpha_1} \left(\mu_0 + \mu_1 z_{t-1} + \mu_2 z_{t-1}^2 - \alpha_0 \right)\right)\right],\label{eq:mod_quad1} \\
\sigma(z_{t-1}) &= \log\left[ 1+ \exp\left( \sigma_0 + \sigma_1 z_{t-1} + \sigma_2 z_{t-1}^2 \right)\right],\label{eq:mod_quad2}
\end{align}
and compare the performance of the same set of variational posteriors as in the benchmark model. In all cases, we optimize the evidence lower bound (ELBO) using reparameterization-based gradient ascent, while also computing indirect variational inference (IVI) estimates using fixed-point iteration. In estimation we impose that both the mean and skewness of $e_t$ and $z_1$ are equal to zero.

\paragraph{Posterior densities.}

\input{figures/hockeystick_posteriors}

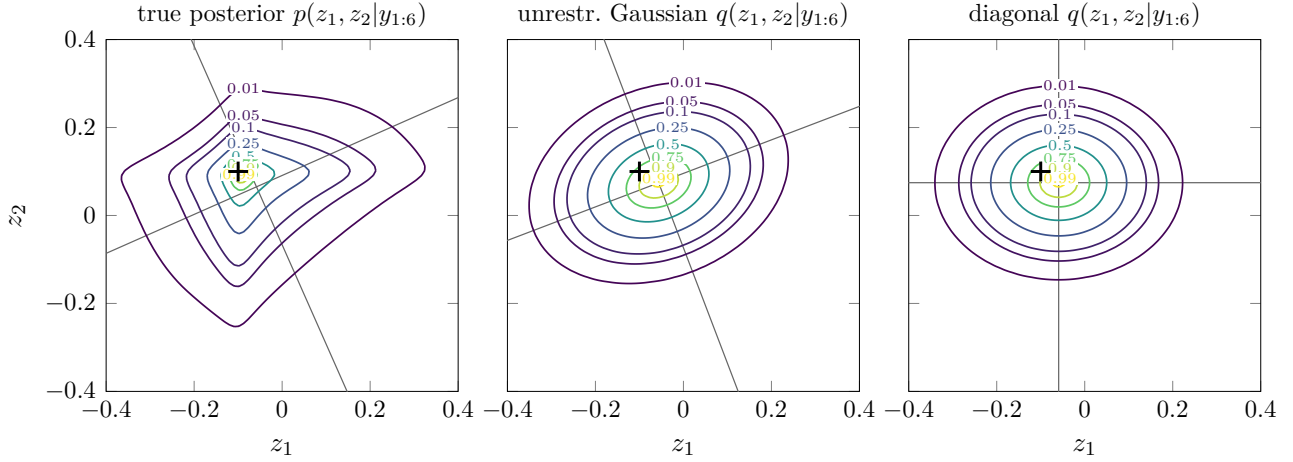
Before turning to the simulation results in the next subsection, it is useful to compare the shape of the variational posterior densities with that of the true posterior. The left panel of Figure~\ref{hockeystick_posteriors} shows contours of the true posterior density $p(z_1,z_2\,|\, y_{1:6})$ implied by the model. For this exercise, we set the parameters to their true values and fix the observations $y_{1},...,y_6$. The posterior density is clearly non-Gaussian, with asymmetric, diamond-shaped contours. 

The middle panel of Figure~\ref{hockeystick_posteriors} shows the contour plots of the best approximating posterior in the Gaussian family with an unrestricted covariance matrix. In this case, contours are ellipsoids. This illustrates that, unlike in the linear model, in the nonlinear model the ``truth-in-class'' assumption is not satisfied because the true posterior does not belong to the Gaussian variational family. At the same time, the unrestricted Gaussian family reproduces the mean (indicated with a cross) and the covariance matrix (indicated by straight lines) quite well. 

In the right panel of Figure~\ref{hockeystick_posteriors} we show the contour plots of the best Gaussian approximating posterior with a diagonal covariance matrix. In this mean-field case, contours are ellipsoids and the mean is well reproduced. However, the covariance structure of the true posterior is not well approximated because the diagonal specification imposes independence between $z_1$ and $z_2$ by construction.

\subsection{Results for the Nonlinear Model}

\input{figures/hockeystick_mu_sigma_z1_e}

Figure~\ref{hockeystick_mu_sigma_z1_e} and the top panel in Table~\ref{hockeystick_parameter_table} summarize the estimation results for the nonlinear specification, for $N=30{,}000$ and $T=6$. As in the linear Gaussian model, using a mean-field variational posterior with a diagonal Gaussian structure or using a model that ignores transitory shocks, we find that the approximation fails to recover the parameters accurately, especially the volatility function $\sigma(\cdot)$ and the density of the transitory shock $e_t$. 

By comparison, the unrestricted Gaussian posterior captures both the central tendency and dispersion of the latent persistent component $z_t$ well, including the curvature in the conditional volatility function. The structured posterior based on a hidden Markov specification performs similarly. However, both fail at approximating the shape of the density of transitory shocks (see the lower right graph in Figure~\ref{hockeystick_mu_sigma_z1_e}). As confirmed by the first and fourth rows of Table~\ref{hockeystick_parameter_table}, these VI approaches tend to severely underestimate the kurtosis of transitory shocks (i.e., they overstate the parameter $\gamma_2$).

\begin{table}[!t]
\centering
\caption{Parameter Estimates in the Nonlinear Model}
\label{hockeystick_parameter_table}
\setlength{\tabcolsep}{4pt}%
\resizebox{\textwidth}{!}{%
\begin{tabular}{lcccccccccccc}
\toprule
 & \multicolumn{5}{c}{$\mu(z_{t-1})$} & \multicolumn{3}{c}{$\sigma(z_{t-1})$} & \multicolumn{2}{c}{$f_\alpha(z_1)$} & \multicolumn{2}{c}{$\psi_\gamma(e_t)$} \\
\cmidrule(l{2pt}r{2pt}){2-6}\cmidrule(l{2pt}r{2pt}){7-9}\cmidrule(l{2pt}r{2pt}){10-11}\cmidrule(l{2pt}r{2pt}){12-13}
Parameter & $\eta_0$ & $\eta_1$ & $\mu_0$ & $\mu_1$ & $\mu_2$ & $\sigma_0$ & $\sigma_1$ & $\sigma_2$ & $\alpha_1$ & $\alpha_2$ & $\gamma_1$ & $\gamma_2$ \\
\midrule
DGP & $-0.25$ & $\phantom{-}0.10$ & $\phantom{-}0.00$ & $\phantom{-}0.90$ & $\phantom{-}0.00$ & $-1.80$ & $\phantom{-}0.00$ & $\phantom{-}0.35$ & $\phantom{-}0.34$ & $\phantom{-}0.89$ & $\phantom{-}0.03$ & $\phantom{-}0.47$ \\
\addlinespace
\multicolumn{13}{l}{\makebox[4.0em][l]{\textbf{$T = 6$}}\textit{Variational posterior}} \\
\hspace*{4.0em}(1) unrestr.\ Gaussian & $-0.28$ & $\phantom{-}0.15$ & $-0.01$ & $\phantom{-}0.95$ & $-0.05$ & $-1.81$ & $\phantom{-}0.05$ & $\phantom{-}0.31$ & $\phantom{-}0.33$ & $\phantom{-}0.88$ & $\phantom{-}0.11$ & $\phantom{-}0.81$ \\
\hspace*{4.0em}(2) tridiagonal & $-0.27$ & $\phantom{-}0.14$ & $-0.01$ & $\phantom{-}0.95$ & $-0.05$ & $-1.79$ & $\phantom{-}0.08$ & $\phantom{-}0.34$ & $\phantom{-}0.34$ & $\phantom{-}0.89$ & $\phantom{-}0.12$ & $\phantom{-}0.84$ \\
\hspace*{4.0em}(3) diagonal & $-0.23$ & $\phantom{-}0.17$ & $-0.04$ & $\phantom{-}0.94$ & $-0.11$ & $-1.42$ & $\phantom{-}0.12$ & $\phantom{-}0.26$ & $\phantom{-}0.36$ & $\phantom{-}0.89$ & $\phantom{-}0.08$ & $\phantom{-}0.99$ \\
\hspace*{4.0em}(4) hidden Markov & $-0.27$ & $\phantom{-}0.15$ & $-0.02$ & $\phantom{-}0.98$ & $-0.06$ & $-1.86$ & $\phantom{-}0.03$ & $\phantom{-}0.39$ & $\phantom{-}0.34$ & $\phantom{-}0.88$ & $\phantom{-}0.11$ & $\phantom{-}0.80$ \\
\addlinespace
\multicolumn{13}{l}{\hspace*{4.0em}\textit{Indirect Variational Inference}} \\
\hspace*{4.0em}(5) unrestr.\ Gaussian (IVI) & $-0.25$ & $\phantom{-}0.10$ & $\phantom{-}0.00$ & $\phantom{-}0.90$ & $\phantom{-}0.00$ & $-1.82$ & $-0.01$ & $\phantom{-}0.36$ & $\phantom{-}0.34$ & $\phantom{-}0.88$ & $\phantom{-}0.03$ & $\phantom{-}0.47$ \\
\addlinespace
\multicolumn{13}{l}{\hspace*{4.0em}\textit{Ignoring transitory shocks}} \\
\hspace*{4.0em}(6) & $-0.22$ & $\phantom{-}0.17$ & $-0.04$ & $\phantom{-}0.90$ & $-0.12$ & $-1.31$ & $\phantom{-}0.13$ & $\phantom{-}0.24$ & $\phantom{-}0.37$ & $\phantom{-}0.90$ & -- & -- \\
\addlinespace
\multicolumn{13}{l}{\makebox[4.0em][l]{\textbf{$T = 40$}}\textit{Variational posterior}} \\
\hspace*{4.0em}(7) unrestr.\ Gaussian & $-0.32$ & $\phantom{-}0.14$ & $\phantom{-}0.00$ & $\phantom{-}0.82$ & $\phantom{-}0.04$ & $-1.60$ & $\phantom{-}0.07$ & $\phantom{-}0.20$ & $\phantom{-}0.34$ & $\phantom{-}0.89$ & $\phantom{-}0.12$ & $\phantom{-}0.92$ \\
\addlinespace
\multicolumn{13}{l}{\hspace*{4.0em}\textit{Indirect Variational Inference}} \\
\hspace*{4.0em}(8) unrestr.\ Gaussian (IVI) & $-0.24$ & $\phantom{-}0.09$ & $\phantom{-}0.00$ & $\phantom{-}0.90$ & $\phantom{-}0.00$ & $-1.81$ & $-0.01$ & $\phantom{-}0.36$ & $\phantom{-}0.32$ & $\phantom{-}0.86$ & $\phantom{-}0.03$ & $\phantom{-}0.47$ \\
\bottomrule
\end{tabular}%
}
\\ \vspace{0.2cm} \figurenote{The table reports the true parameter values of the simulated nonlinear DGP alongside estimates at two panel lengths, $T=6$ (top panel) and $T=40$ (bottom panel), for different variational-posterior specifications, the IVI correction of the unrestricted-Gaussian posterior, and a specification that ignores transitory shocks (only for $T=6$). The estimation model follows equations (\ref{eq:mod_quad1})--(\ref{eq:mod_quad2}), applied to data generated with sinh-arcsinh innovations for $z_1$ and $e_t$. Nonlinear DGP, $N=30{,}000$.}
\end{table}

This mixed evidence motivates relying on the variational specification as an auxiliary model through indirect inference. In the fifth row of Table~\ref{hockeystick_parameter_table} we show IVI parameter estimates obtained using fixed-point iteration, with $\kappa=0.6$ and 25 iterations, starting from the unrestricted Gaussian variational estimates. The IVI method succeeds at correcting biases throughout: parameters obtained using indirect variational inference are very close to the true values. In particular, IVI accurately captures the excess kurtosis in transitory shocks.

\paragraph{Performance for longer $T$.} 

The scalability of variational inference is an important potential advantage relative to other methods. This feature gives the possibility to entertain settings that would otherwise be too computationally demanding. To evaluate the behavior of VI and IVI in larger data sets, in the bottom panel of Table~\ref{hockeystick_parameter_table} we report results based on a panel with $T=40$ periods, keeping the cross-sectional size constant ($N=30{,}000$). The estimates show that VI, based on an unrestricted Gaussian specification, exhibits some biases, most notably by understating the kurtosis of transitory shocks (as shown by a too large estimate of $\gamma_2$). However, IVI corrects most of the bias and delivers good performance across parameters. This is a promising finding for the use of VI and IVI in large data sets such as administrative earnings records. Appendix \ref{app:resources} provides additional information about computational and memory cost of VI across sample sizes relative to Sequential Monte Carlo.

\section{Extensions: Heterogeneity and Serial Correlation}\label{sec:MA1}

In this section, we show how the variational approach can be easily modified to handle two important extensions of the baseline nonlinear model incorporating time-invariant heterogeneity and serially correlated transitory shocks.

\subsection{Time-Invariant Heterogeneity\label{sec:hetero-model}}

We first augment the nonlinear model to allow for a time-invariant latent type $a$. We postulate model (\ref{eq:mod_1_het})-(\ref{eq:mod_2_het})-(\ref{eq:mod_3_het}) that features a latent time-invariant type $a$. Let $\theta=(\mu(\cdot,\cdot),\sigma(\cdot,\cdot),\alpha)$, and let $f_{\theta}$ denote the joint density of $a,z_1,...,z_T$. As before, let $\vartheta=(\theta,\gamma)$. The log-likelihood function for a single individual is
\begin{align}
{\cal{L}}_{\vartheta}(y_{1:T}) = \log \int f_\theta(a,z_{1:T}) \, \psi_\gamma(y_{1:T} - z_{1:T};a) \, dz_{1:T} \, da,\label{eq:lik_het}
\end{align}
where now the integral is taken with respect to $(a,z_1,...,z_T)$. Given a variational posterior density
$q_\phi(a,z_{1:T} \,|\, y_{1:T})$, the ELBO is given by
\begin{equation}
 \mathcal{E}_{\vartheta,\phi}(y_{1:T}) 
= \mathbb{E}_{q_\phi(a,z_{1:T} \,|\, y_{1:T})}\left[\log \frac{f_\theta(a,z_{1:T})  \psi_\gamma(y_{1:T} - z_{1:T};a)}{q_\phi(a,z_{1:T} \,|\, y_{1:T})}\right].\label{eq:ELBO_het}
\end{equation}

\paragraph{Parameterization.} 
We evaluate the performance of VI and IVI in a nonlinear model where the variance of transitory shocks is heterogeneous across individuals, as in \citet{almuzara2020heterogeneity}:
\begin{align}
&y_t = z_t + e^{a} \cdot  e_t, \\
&z_t = \mu(z_{t-1}) + \sigma(z_{t-1}) u_t.
\end{align}
As before, $e_t$ and $z_1$ follow (\ref{eq:sinh-e}) and (\ref{eq:sinh-z1}) with $(\gamma_1,\gamma_2) = (1,0.47)$ and $(\alpha_1,\alpha_2) = (0.34,0.89)$, respectively, and $u_t \sim \mathcal{N}(0,1)$, while both the mean and skewness of $e_t$ and $z_1$ are set to zero. Here $\gamma_1 = 1$ is a normalization. We use 
\begin{align}
\mu(z_{t-1}) & =  \mu_0 + \mu_1 z_{t-1} + \mu_2 z_{t-1}^2 , \\
\sigma(z_{t-1}) &=  \log\left(1+ \exp\left( \sigma_0 +\sigma_1 z_{t-1} + \sigma_2 z_{t-1}^2 \right)\right), \\
a|z_1 &\sim \mathcal{N}(\lambda_0 + \lambda_1 z_1,\sigma_a^2),
\end{align}
where $(\mu_0,\mu_1,\mu_2) = (0,0.9,0)$, $(\sigma_0,\sigma_1,\sigma_2) = (-1.8,0,0.35)$, and $(\lambda_0,\lambda_1,\sigma_a) = (-2.3, 0.26, 0.29)$.

\paragraph{Results.}

In Figure~\ref{hetero_scale_a_sigma_z1_e} and Table~\ref{hetero_scale_parameter_table} we report estimates based on two Gaussian VI specifications --- with an unrestricted and diagonal covariance matrix, respectively -- as well as IVI estimates. While the diagonal specification overestimates the volatility $\sigma(z_{t-1})$, VI with an unrestricted covariance underestimates it. Moreover, both the diagonal and unrestricted Gaussian VI methods tend to underestimate the amount of variance heterogeneity. By comparison, IVI corrects these biases and is overall close to the true parameter values.

\input{figures/hetero_scale_a_sigma_z1_e}

\begin{table}[!t]
\centering
\caption{Parameter Estimates in the Heterogeneity Model}
\label{hetero_scale_parameter_table}
\setlength{\tabcolsep}{4pt}%
\resizebox{\textwidth}{!}{%
\begin{tabular}{lcccccccccccc}
\toprule
 & \multicolumn{3}{c}{$\mu(z_{t-1})$} & \multicolumn{3}{c}{$\sigma(z_{t-1})$} & \multicolumn{2}{c}{$f_\alpha(z_1)$} & \multicolumn{3}{c}{$p(a\,|\,z_1)$} & \multicolumn{1}{c}{$\psi_\gamma(e_t)$} \\
\cmidrule(l{2pt}r{2pt}){2-4}\cmidrule(l{2pt}r{2pt}){5-7}\cmidrule(l{2pt}r{2pt}){8-9}\cmidrule(l{2pt}r{2pt}){10-12}\cmidrule(l{2pt}r{2pt}){13-13}
Parameter & $\mu_0$ & $\mu_1$ & $\mu_2$ & $\sigma_0$ & $\sigma_1$ & $\sigma_2$ & $\alpha_1$ & $\alpha_2$ & $\lambda_0$ & $\lambda_1$ & $\sigma_a$ & $\gamma_2$ \\
\midrule
DGP & $\phantom{-}0.00$ & $\phantom{-}0.90$ & $\phantom{-}0.00$ & $-1.80$ & $\phantom{-}0.00$ & $\phantom{-}0.35$ & $\phantom{-}0.34$ & $\phantom{-}0.89$ & $-2.30$ & $\phantom{-}0.26$ & $\phantom{-}0.29$ & $\phantom{-}0.47$ \\
\addlinespace
\multicolumn{13}{l}{\textit{Variational posterior}} \\
(1) unrestr.\ Gaussian & $\phantom{-}0.00$ & $\phantom{-}0.94$ & $\phantom{-}0.00$ & $-2.12$ & $\phantom{-}0.02$ & $\phantom{-}0.07$ & $\phantom{-}0.33$ & $\phantom{-}0.91$ & $-1.80$ & $\phantom{-}0.21$ & $\phantom{-}0.23$ & $\phantom{-}0.56$ \\
(2) diagonal & $\phantom{-}0.00$ & $\phantom{-}0.63$ & $-0.02$ & $-0.99$ & $\phantom{-}0.05$ & $\phantom{-}0.16$ & $\phantom{-}0.35$ & $\phantom{-}0.86$ & $-2.13$ & $\phantom{-}0.03$ & $\phantom{-}0.13$ & $\phantom{-}0.52$ \\
\addlinespace
\multicolumn{13}{l}{\textit{Indirect Variational Inference}} \\
(3) unrestr.\ Gaussian (IVI) & $\phantom{-}0.00$ & $\phantom{-}0.91$ & $\phantom{-}0.00$ & $-1.86$ & $\phantom{-}0.07$ & $\phantom{-}0.34$ & $\phantom{-}0.31$ & $\phantom{-}0.85$ & $-2.22$ & $\phantom{-}0.24$ & $\phantom{-}0.26$ & $\phantom{-}0.48$ \\
\bottomrule
\end{tabular}%
}
\\ \vspace{0.2cm} \figurenote{True parameter values of the simulated heterogeneity-model DGP and the estimates under two variational posteriors and the IVI correction of the unrestricted Gaussian posterior. The transitory shock has a fixed baseline dispersion ($\gamma_1=1$) that each individual rescales by $e^{a}$, with $a|z_1\sim\mathcal N(\lambda_0+\lambda_1 z_1,\sigma_a^2)$. Nonlinear DGP with heterogeneity, $N=30{,}000$, $T=6$.}
\end{table}

\subsection{Serially Correlated Transitory Shocks\label{subsec_MA1}}

We then show how to extend the earnings dynamics model to allow for serial correlation in the transitory shock $e_t$. Allowing for an MA(1) component in the transitory shock directly relates to the empirical strategy of \citet{MeghirPistaferri2004}, who emphasize the importance of serial correlation in transitory shocks for accurately characterizing the dynamics of income volatility. Specifically, we introduce an MA(1) structure in the transitory component \( e_t \), while maintaining the Markovian evolution of the latent persistent component \( z_t \). The data-generating process is specified as:
\begin{align}
y_t &= z_t + e_t, \\
z_t &= \mu( z_{t-1}) + \sigma(z_{t-1}) u_t, \\
e_t &= \varepsilon_t + \zeta \varepsilon_{t-1}, \\
z_1 &\sim f_\alpha, \quad u_t \sim \mathcal{N}(0,1), \quad \varepsilon_t \sim \psi_{\gamma}.
\end{align}

The new feature is the presence of serial dependence in the transitory shock process: each observed income \( y_t \) now depends not only on the current latent state \( z_t \), but also indirectly on the past component \( \varepsilon_{t-1} \). As a result, \( y_t \) remains serially correlated even after conditioning on the latent state \( z_{t} \), and the likelihood function no longer factorizes across time conditional on the latent states. Moreover, $\varepsilon_{t-1}$ is only partially revealed by \(y_{t-1}\), which itself contains \(\varepsilon_{t-2}\). Consequently, the density $p(z_t \,|\, z_{t-1}, y_{1:T})$ generally depends on a backward window of past observations \(y_{t-L:t-1}\) with \(L \geq 1\), together with future observations \(y_{t:T}\). 

The empirical evidence in the PSID in the next section is based on a model with first-order moving average transitory shocks, suitably augmented to also allow for individual-specific transitory variances, and estimation relies on a Gaussian variational family with an unrestricted covariance matrix. In Appendix \ref{app:MA1} we describe an alternative approach that imposes the model's dynamic structure on the variational family.

\section{Nonlinear Earnings Processes in the PSID}\label{sec:PSID}

In this section we apply variational inference to real-world earnings data. Our objective is to compare the earnings process estimated using variational inference to the one reported in \citet{ArellanoBlundellBonhomme2017}, who estimate nonlinear earnings processes using a quantile-based stochastic EM approach. Relative to this paper, we rely on a specification for annual income (as opposed to biennial) with heterogeneous transitory variances and serially correlated transitory shocks, and estimate the model using variational methods.

\subsection{Estimated Earnings Process}

\paragraph{A nonlinear process with heterogeneous and serially correlated transitory shocks.}

The model we estimate combines the various ingredients that we have introduced in the previous sections. The full specification is as follows:
\begin{align}
y_t &= z_t + e^{a} \cdot e_t, \\
z_t &= \mu( z_{t-1}) + \sigma(z_{t-1}) u_t,  \quad u_t \sim \mathcal{N}(0,1), \\
e_t &= \varepsilon_t + \zeta \varepsilon_{t-1}, \quad
\varepsilon_t \sim \Phi\!\left( \sinh\!\big( \gamma_2\,\operatorname{asinh}(\varepsilon -\gamma_0) -\gamma_3 \big) \right),\\
a|z_1 &\sim \mathcal{N}(\lambda_0 + \lambda_1 z_1,\sigma_a^2), \quad z_1 \sim \Phi\!\left( \sinh\!\left( \alpha_2\,\operatorname{asinh}\left(\tfrac{z-\alpha_0}{\alpha_1}\right) - \alpha_3\right) \right), \\
\mu(z_{t-1}) & =  \mu_0 + \mu_1 z_{t-1} + \mu_2 z_{t-1}^2 , \\
\sigma(z_{t-1}) &=  \log\left(1+ \exp\left( \sigma_0 +\sigma_1 z_{t-1} + \sigma_2 z_{t-1}^2 \right)\right),
\end{align}
where the transitory innovations $\varepsilon_{t}$ and initial conditions $z_1$ can exhibit skewness since we do not impose that $\gamma_3=0$ or $\alpha_3=0$. The parameters $(\gamma_0,\alpha_0)$ are mean-centering terms that ensure that $\varepsilon_t$ and $z_1$ have zero mean. The transitory shocks $e^{a} e_t$ have heterogeneous variances, since the latent type $a$ differs across individuals, and they are serially correlated whenever $\zeta\neq 0$.

\input{figures/psid_a_sigma_z1_eps}

We estimate this model using data from the Panel Study of Income Dynamics (PSID) covering the period 1980--1989. The PSID is a long-running U.S. household survey with detailed annual information on household labor earnings, hours worked, employment status, and demographic characteristics. Following \citet{BlundellEtAl2008}, we restrict the sample to male household heads aged 25--60 with strong labor force attachment, excluding the self-employed, individuals reporting fewer than 520 annual hours, and those with missing or implausible income reports. Household labor income is constructed net of transfers and taxes using the PSID family files.\footnote{We apply the procedure from the replication package in \citet{BlundellEtAl2008}, where log household earnings are residualized on a set of demographic variables in a first step.} The final sample is a balanced panel of $N=741$ households observed annually for $T=10$ years.

\begin{table}[!t]
\centering
\caption{Estimates on the PSID}
\label{psid_parameter_table}
\setlength{\tabcolsep}{3pt}%
\resizebox{\textwidth}{!}{%
\begin{tabular}{lccccccccccccccc}
\toprule
 & \multicolumn{3}{c}{$\mu(z_{t-1})$} & \multicolumn{3}{c}{$\sigma(z_{t-1})$} & \multicolumn{3}{c}{$f_\alpha(z_1)$} & \multicolumn{3}{c}{$p(a\,|\,z_1)$} & \multicolumn{2}{c}{$\psi_\gamma(\varepsilon)$} & \multicolumn{1}{c}{$\mathrm{MA}(1)$} \\
\cmidrule(l{2pt}r{2pt}){2-4}\cmidrule(l{2pt}r{2pt}){5-7}\cmidrule(l{2pt}r{2pt}){8-10}\cmidrule(l{2pt}r{2pt}){11-13}\cmidrule(l{2pt}r{2pt}){14-15}\cmidrule(l{2pt}r{2pt}){16-16}
Parameter & $\mu_0$ & $\mu_1$ & $\mu_2$ & $\sigma_0$ & $\sigma_1$ & $\sigma_2$ & $\alpha_1$ & $\alpha_2$ & $\alpha_3$ & $\lambda_0$ & $\lambda_1$ & $\sigma_a$ & $\gamma_2$ & $\gamma_3$ & $\zeta$ \\
\midrule
\multicolumn{16}{l}{\textit{Variational posterior}} \\
(1) unrestr.\ Gaussian & $\phantom{-}0.00$ & $\phantom{-}1.01$ & $\phantom{-}0.02$ & $-2.86$ & $-0.05$ & $\phantom{-}0.13$ & $\phantom{-}0.20$ & $\phantom{-}0.73$ & $-0.08$ & $-2.13$ & $-0.20$ & $\phantom{-}0.40$ & $\phantom{-}0.79$ & $-0.12$ & $\phantom{-}0.35$ \\
(2) diagonal & $-0.01$ & $\phantom{-}0.94$ & $\phantom{-}0.03$ & $-1.77$ & $-0.28$ & $\phantom{-}0.48$ & $\phantom{-}0.18$ & $\phantom{-}0.65$ & $-0.09$ & $-2.66$ & $-0.04$ & $\phantom{-}0.11$ & $\phantom{-}0.77$ & $\phantom{-}0.01$ & $-0.26$ \\
\addlinespace
\multicolumn{16}{l}{\textit{Indirect Variational Inference}} \\
(3) unrestr.\ Gaussian (IVI) & $\phantom{-}0.00$ & $\phantom{-}1.00$ & $\phantom{-}0.03$ & $-2.44$ & $-0.14$ & $\phantom{-}0.75$ & $\phantom{-}0.18$ & $\phantom{-}0.68$ & $-0.11$ & $-2.83$ & $-0.18$ & $\phantom{-}0.58$ & $\phantom{-}0.62$ & $-0.14$ & $\phantom{-}0.26$ \\[-0.6ex]
 & {\footnotesize $\phantom{-}(0.00)$} & {\footnotesize $\phantom{-}(0.01)$} & {\footnotesize $\phantom{-}(0.01)$} & {\footnotesize $\phantom{-}(0.09)$} & {\footnotesize $\phantom{-}(0.17)$} & {\footnotesize $\phantom{-}(0.17)$} & {\footnotesize $\phantom{-}(0.04)$} & {\footnotesize $\phantom{-}(0.09)$} & {\footnotesize $\phantom{-}(0.06)$} & {\footnotesize $\phantom{-}(0.15)$} & {\footnotesize $\phantom{-}(0.18)$} & {\footnotesize $\phantom{-}(0.04)$} & {\footnotesize $\phantom{-}(0.04)$} & {\footnotesize $\phantom{-}(0.03)$} & {\footnotesize $\phantom{-}(0.03)$} \\
\bottomrule
\end{tabular}%
}
\\ \vspace{0.2cm} \figurenote{The table reports parameter estimates for the nonlinear earnings model on the PSID, 1980--1989 ($N=741$, $T=10$), under two variational-posterior specifications (unrestricted Gaussian, diagonal) and the IVI correction of the unrestricted-Gaussian posterior. Reported parameters are those of the conditional mean $\mu(z_{t-1})$, conditional volatility $\sigma(z_{t-1})$, the initial state $z_1$, the individual scale heterogeneity $p(a\,|\,z_1)$, and the transitory shock $\varepsilon_t$ with its MA(1) coefficient $\zeta$. Standard errors in parentheses are from a nonparametric bootstrap that resamples households (100 replications).}
\end{table}

Figure~\ref{psid_a_sigma_z1_eps} and Table~\ref{psid_parameter_table} show the estimates based on two Gaussian VI specifications (with diagonal and unrestricted covariance, respectively), and IVI estimates obtained by fixed-point iteration starting from the unrestricted Gaussian VI estimates. Our IVI estimates reveal a highly persistent latent earnings process. The conditional mean $\mu(z_{t-1})$ is close to linear, with an autoregressive coefficient approximately equal to one. The conditional volatility $\sigma(z_{t-1})$ exhibits a U shape: volatility is high at the lower end of the distribution, decreases toward the middle, and rises again at the upper end. In addition, we find substantial heterogeneity in transitory variances, as well as positive serial correlation in transitory shocks with a coefficient of 0.26. Confidence bands (obtained by bootstrap clustered at the household level) show that estimates are relatively precise despite the small sample size. The IVI estimates differ from the VI estimates, in line with our evidence on simulated data. The main differences are for the curvature of the volatility, which VI understates relative to IVI, and for the amount of variance heterogeneity, which VI underestimates as well. 

Our IVI estimates imply that persistence in $z_t$ is nonlinear. As in \citet{ArellanoBlundellBonhomme2017}, we measure persistence as the derivative of the conditional quantile function of $z_t$ given $z_{t-1}$ with respect to $z_{t-1}$. In words, persistence measures how the current earnings component $z_t$ changes when the past component $z_{t-1}$ changes, for given values of the latter and the shock $u_t$. In our conditionally Gaussian model the quantile function is
$$Q_{\tau}(z_t\,|\, z_{t-1})=\mu(z_{t-1})+\sigma(z_{t-1})\Phi^{-1}(\tau),\quad \text{for all }\tau\in(0,1),$$
with $\Phi$ the standard normal cdf. We then measure persistence as
\begin{align}
\rho(z_{t-1},\tau)&=\nabla_{z_{t-1}} Q_{\tau}(z_t\,|\, z_{t-1})\notag\\
&=\underset{\text{state-dependent mean}}{\underbrace{\nabla_{z_{t-1}}\mu(z_{t-1})}}+\underset{\text{state-dependent volatility}}{\underbrace{\nabla_{z_{t-1}}\sigma(z_{t-1})\Phi^{-1}(\tau)}}.\label{eq:rho}
\end{align}
According to our estimates, the state-dependent mean component is close to a constant as $\mu(z_{t-1})$ is approximately linear. In contrast, the state-dependent volatility component is U-shaped in $z_{t-1}$, which implies that the persistence measure $\rho(z_{t-1},\tau)$ depends both on the state $z_{t-1}$ and the shock $\tau$ (i.e., the percentile rank of $u_t$).

\input{figures/psid_persistence_residual}

We report our IVI estimate of the persistence surface in Panel (a) of Figure~\ref{psid_persistence_residual}. The two horizontal axes indicate the percentile rank of $z_{t-1}$ and the percentile rank $\tau$ of $u_t$, respectively, and the vertical axis indicates the values of $\rho(z_{t-1},\tau)$. We see that persistence is approximately constant, and close to $\rho=1$, for central values of $z_{t-1}$ and $u_t$. However, high-$u_t$ shocks for low-$z_{t-1}$ households are associated with a lower persistence, as low as 0.5. We also observe that low-$u_t$ shocks for high-$z_{t-1}$ households are also associated with a lower $\rho$, although the decrease is not as stark. Lastly, our estimates indicate some increase in $\rho$ for high-$u_t$/high-$z_{t-1}$ and low-$u_t$/low-$z_{t-1}$ combinations.\footnote{While the persistence surface is not identical to the one in \citet{ArellanoBlundellBonhomme2017}, due to different specifications, estimation methods, and samples (\citealp{ArellanoBlundellBonhomme2017} use biennial post-1999 PSID data) the two surfaces tend to agree.} 

In Panel (b) of Figure~\ref{psid_persistence_residual}, we report the estimated density of the scaled transitory shocks $e^a e_t$. In the model, transitory shocks have heterogeneous variances, and as a result the density estimated by IVI shows a large amount of excess kurtosis relative to the Gaussian. Moreover, as the comparison with the density of $\mathbb{E}[e^a]e_t$ shows, only about half of the deviation from the Gaussian is explained by the variance heterogeneity, the other half coming from the fact that the shocks $e_t$ themselves are non-Gaussian.

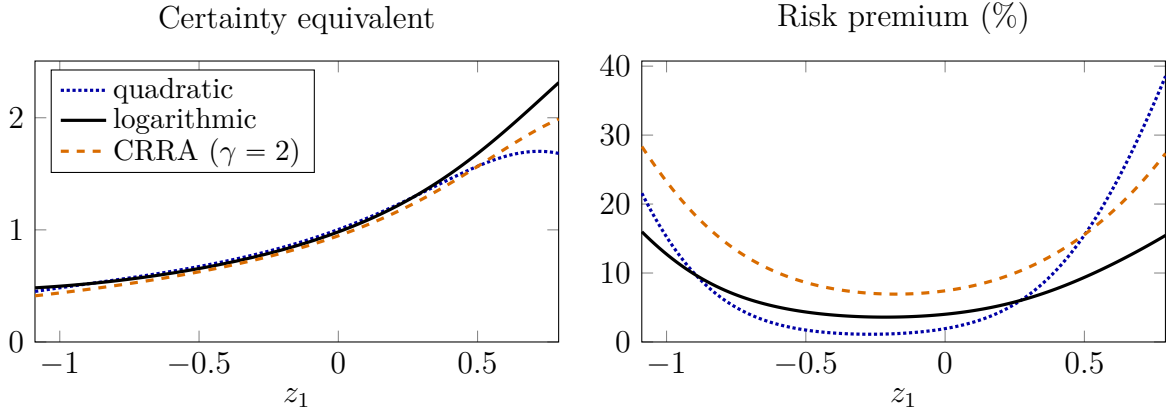
\subsection{Certainty Equivalent and Risk Premium}

To assess the economic relevance of the features of the earnings process that we have estimated, we next report several summaries that quantify (under certain assumptions) the risk faced by households. To proceed, let \( z_t \sim f_t(z_t \,|\, z_1) \) denote the distribution of future outcomes conditional on the initial latent state \( z_1 \), and let \( \beta \in (0,1) \) denote the discount factor. Let $u(z_t)=U(e^{z_t})$ denote household utility, where the utility function $u$ does not vary over time. Here we interpret the persistent earnings component $z_t$ as a proxy for log consumption.

The household's expected utility is given by
\begin{equation}
\mathbb{E} \left[ \sum_{t=1}^\infty \beta^{t-1} u(z_t) \right]
= \sum_{t=1}^\infty \beta^{t-1} \int u(z) f_t(z \,|\, z_1) \, dz.
\end{equation}
Rewriting this expression by exchanging the order of summation and integration yields:
\begin{equation}
\mathbb{E} \left[ \sum_{t=1}^\infty \beta^{t-1} u(z_t) \right]
= \int u(z) \left( \sum_{t=1}^\infty \beta^{t-1} f_t(z \,|\, z_1) \right) dz.
\end{equation}
We define the \textit{discounted mixture density} as
\begin{equation}\label{eq:ftilde}
\widetilde{f}(z \,|\, z_1) = (1 - \beta) \sum_{t=1}^\infty \beta^{t-1} f_t(z \,|\, z_1),
\end{equation}
which integrates to one and thus defines a valid probability distribution.

\input{figures/psid_mixture_density}

Given $\widetilde{f}(z \,|\, z_1)$, and for any continuous, integrable utility function $U(\cdot)$, expected discounted sums of utility can be computed in closed form as a single integral:
\begin{equation}
\mathbb{E} \left[ \sum_{t=1}^\infty \beta^{t-1} u(z_t) \right]
= \frac{1}{1 - \beta} \int u(z) \widetilde{f}(z \,|\, z_1) \, dz
= \frac{1}{1 - \beta} \mathbb{E}_{\widetilde{f}}[u(z)].
\end{equation}
For example, the \textit{certainty equivalent} \( c^{CE} \) defined as the solution to
\begin{equation}
\sum_{t=1}^\infty \beta^{t-1} U(c^{CE}) = \mathbb{E} \left[ \sum_{t=1}^\infty \beta^{t-1} u(z_t) \right],
\end{equation}
can be obtained as
\begin{equation}
c^{CE} = U^{-1} \left( \mathbb{E}_{\widetilde{f}}\!\left[ U(e^z) \right] \right).
\end{equation}
In turn, the \emph{risk premium} -- defined as the amount an agent would be willing to pay to avoid uncertainty -- can be obtained as follows:
\begin{equation}
\pi = 1 - \dfrac{c^{CE}}{\mathbb{E}_{\widetilde{f}}[e^z]}.
\end{equation}



In Figure~\ref{psid_mixture_density} we report estimates of the discounted density $\widetilde{f}$ estimated on the PSID data, for various values of $z_1$, net of the point mass at $z_1$. The conditional dispersion of discounted log earnings is smaller for average values of $z_1$ and larger for high and low $z_1$, reflecting the shape of the volatility function $\sigma(\cdot)$. In addition, the conditional densities are skewed to the right.

\input{figures/psid_ce_risk_premium}

In Figure~\ref{psid_ce_risk_premium} we report certainty equivalents and risk premia estimated from the PSID for $\beta=0.9$ under quadratic, logarithmic, and CRRA utility (with parameter $2.0$). The certainty equivalent is increasing and convex in the initial latent state, reaching values at the top of the distribution that are four to five times higher than those at the bottom. In addition, we find that risk premia display substantial heterogeneity across the distribution. For example, under log utility they range from about 5 percent in the middle of the distribution to 10 percent at the lower end. With more curvature in preferences, as under CRRA utility, risk premia become even larger, approaching 20 percent for households starting at either tail of the distribution.

\section{Conclusion}\label{sec:conclusion}

While a growing body of evidence highlights the relevance of nonlinear features in earnings dynamics, nonlinear state-space models remain challenging to estimate. In this paper we explore whether variational inference (VI) can provide a reliable alternative to existing methods. We propose a flexible framework that nests the canonical linear Gaussian process while accommodating nonlinear persistence, state-dependent volatility, serially correlated or heavy-tailed transitory shocks, and latent time-invariant heterogeneity. We rely on variational approximations to the posterior distribution of latent states, based on Gaussian specifications, to estimate various versions of the model.

Our simulation results show that the bias of variational inference can be large, in particular when using a mean-field specification in a dynamic setting. However, choosing a sufficiently flexible family, and using it as an auxiliary model through indirect variational inference (IVI), provides successful strategies in our simulations. Applying VI and IVI to PSID data, we find a nearly linear conditional mean (with average persistence close to unity), a U-shaped conditional variance across the income distribution, and evidence of variance heterogeneity and serial correlation in transitory shocks. As in \citet{ArellanoBlundellBonhomme2017}, persistence is lower for high-earnings households experiencing negative shocks and low-earnings households experiencing positive shocks.

Taken together, these findings motivate further study of VI and IVI as tractable alternatives to traditional likelihood-based methods to estimate realistic models of earnings dynamics. The approach scales well to long panels, is compatible with modern optimization frameworks, and retains flexibility to model key nonlinearities that matter empirically. These features make it feasible to estimate flexible nonlinear models not only in survey data such as the PSID but also in large-scale administrative datasets with richer earnings histories. Generative model estimation is an active area in machine learning, including VI but also Wasserstein methods (\citealp{bernton2017inference}) and Generative Adversarial Networks (e.g., \citealp{kaji2023adversarial}). It is our hope that these methods can help uncover new empirical patterns of dynamics and heterogeneity in earnings data.

\newpage

\bibliographystyle{apalike}
\bibliography{ref}

\clearpage
\appendix

\section{Asymptotic Variances\label{app:var}}

\subsection{Expression of $V^{\rm VI}$}

We follow the approach in \citet{westling2019beyond} and assume their conditions are satisfied. By (\ref{eq:ELBO}) and the envelope theorem, $\widehat{\vartheta}^{\rm VI}$ satisfies $\widehat s(\widehat{\vartheta}^{\rm VI})=0$, where
\begin{equation}
\widehat s(\vartheta)=\widehat{\mathbb{E}}\left[\mathbb{E}_{q_{\widehat\phi(\vartheta)}(z_{1:T} \,|\, y_{1:T})}\left[\nabla_{\vartheta}\log \left(f_\theta(z_{1:T})  \psi_\gamma(y_{1:T}- z_{1:T})\right)\right]\right],
\end{equation}
for
\begin{equation}
\widehat\phi(\vartheta)=\underset{\phi}{\mbox{argmax}}\,\widehat{\mathbb{E}}\left[\mathbb{E}_{q_\phi(z_{1:T} \,|\, y_{1:T})}\left[\log \frac{f_\theta(z_{1:T})  \psi_\gamma(y_{1:T} - z_{1:T})}{q_\phi(z_{1:T} \,|\, y_{1:T})}\right]\right].
\end{equation}
Let 
\begin{equation}
\overline\phi_0(\vartheta)=\underset{\phi}{\mbox{argmax}}\,{\mathbb{E}}_{\vartheta_0}\left[\mathbb{E}_{q_\phi(z_{1:T} \,|\, y_{1:T})}\left[\log \frac{f_\theta(z_{1:T})  \psi_\gamma(y_{1:T} - z_{1:T})}{q_\phi(z_{1:T} \,|\, y_{1:T})}\right]\right].
\end{equation}
In addition, write
\begin{equation}
\widehat s(\vartheta)=\widehat{\mathbb{E}}\left[\nabla_{\vartheta} m\left(\vartheta,\widehat\phi(\vartheta),y_{1:T}\right)\right],
\end{equation}
for
$$ m\left(\vartheta,\phi,y_{1:T}\right)=\mathbb{E}_{q_\phi(z_{1:T} \,|\, y_{1:T})}\left[\log \frac{f_\theta(z_{1:T})  \psi_\gamma(y_{1:T} - z_{1:T})}{q_\phi(z_{1:T} \,|\, y_{1:T})}\right].$$
Taylor expanding, we have
\begin{align*}
\widehat{\mathbb{E}}\left[\nabla_{\vartheta} m\left(\vartheta,\widehat\phi(\vartheta),y_{1:T}\right)\right]&=\widehat{\mathbb{E}}\left[\nabla_{\vartheta} m\left(\vartheta,\overline\phi_0(\vartheta),y_{1:T}\right)\right]\\&+\widehat{\mathbb{E}}\left[\nabla_{\vartheta\phi} m\left(\vartheta,\overline\phi_0(\vartheta),y_{1:T}\right)\right]\left(\widehat\phi(\vartheta)-\overline\phi_0(\vartheta)\right)+o_p(N^{-\frac{1}{2}}).
\end{align*}
Now,
$$\widehat{\mathbb{E}}\left[\nabla_{\phi} m\left(\vartheta,\widehat\phi(\vartheta),y_{1:T}\right)\right]=0,$$
so
$$\widehat\phi(\vartheta)-\overline\phi_0(\vartheta)=-\left(\widehat{\mathbb{E}}\left[\nabla_{\phi\phi} m\left(\vartheta,\overline\phi_0(\vartheta),y_{1:T}\right)\right]\right)^{-1}\widehat{\mathbb{E}}\left[\nabla_{\phi} m\left(\vartheta,\overline\phi_0(\vartheta),y_{1:T}\right)\right]+o_p(N^{-\frac{1}{2}}).$$
Hence
\begin{align*}
&\widehat s(\vartheta)=\widehat{\mathbb{E}}\left[\nabla_{\vartheta} m\left(\vartheta,\widehat\phi(\vartheta),y_{1:T}\right)\right]\\&=\widehat{\mathbb{E}}\left[\nabla_{\vartheta} m\left(\vartheta,\overline\phi_0(\vartheta),y_{1:T}\right)\right]\\&-{\mathbb{E}}_{\vartheta_0}\left[\nabla_{\vartheta\phi} m\left(\vartheta,\overline\phi_0(\vartheta),y_{1:T}\right)\right]\left({\mathbb{E}}_{\vartheta_0}\left[\nabla_{\phi\phi} m\left(\vartheta,\overline\phi_0(\vartheta),y_{1:T}\right)\right]\right)^{-1}\widehat{\mathbb{E}}\left[\nabla_{\phi} m\left(\vartheta,\overline\phi_0(\vartheta),y_{1:T}\right)\right]+o_p(N^{-\frac{1}{2}}).
\end{align*}
Evaluating the score expansion at $\vartheta=\overline\vartheta_0$ and defining the {orthogonalized score}
\begin{equation}
\widetilde s\left(y_{1:T}\right)=\nabla_{\vartheta} m\left(\overline\vartheta_0,\overline\phi_0,y_{1:T}\right)-M_{\vartheta\phi}\,M_{\phi\phi}^{-1}\,\nabla_{\phi} m\left(\overline\vartheta_0,\overline\phi_0,y_{1:T}\right),
\end{equation}
where $\overline\phi_0=\overline\phi_0(\overline\vartheta_0)$ and
\begin{equation}
M_{\vartheta\phi}={\mathbb{E}}_{\vartheta_0}\left[\nabla_{\vartheta\phi} m\left(\overline\vartheta_0,\overline\phi_0,y_{1:T}\right)\right],\qquad
M_{\phi\phi}={\mathbb{E}}_{\vartheta_0}\left[\nabla_{\phi\phi} m\left(\overline\vartheta_0,\overline\phi_0,y_{1:T}\right)\right],
\end{equation}
the expansion of the score reads
\begin{equation}
\widehat s(\overline\vartheta_0)=\widehat{\mathbb{E}}\left[\widetilde s\left(y_{1:T}\right)\right]+o_p(N^{-\frac{1}{2}}).
\end{equation}
Since ${\mathbb{E}}_{\vartheta_0}\left[\nabla_{\vartheta} m\left(\overline\vartheta_0,\overline\phi_0,y_{1:T}\right)\right]=0$ and ${\mathbb{E}}_{\vartheta_0}\left[\nabla_{\phi} m\left(\overline\vartheta_0,\overline\phi_0,y_{1:T}\right)\right]=0$ by the population first-order conditions, we have ${\mathbb{E}}_{\vartheta_0}\left[\widetilde s\left(y_{1:T}\right)\right]=0$, and by the central limit theorem
\begin{equation}
\sqrt{N}\,\widehat s(\overline\vartheta_0)\overset{d}{\rightarrow}{\cal{N}}\left(0,\widetilde\Omega\right),\qquad \widetilde\Omega={\mathbb{E}}_{\vartheta_0}\left[\widetilde s\left(y_{1:T}\right)\widetilde s\left(y_{1:T}\right)^\top\right].
\end{equation}
For the Jacobian, differentiating $\widehat s(\vartheta)=\widehat{\mathbb{E}}\left[\nabla_{\vartheta} m\left(\vartheta,\widehat\phi(\vartheta),y_{1:T}\right)\right]$ totally in $\vartheta$ and using the implicit-function expression
\begin{equation}
\nabla_{\vartheta}\widehat\phi(\vartheta)=-\left(\widehat{\mathbb{E}}\left[\nabla_{\phi\phi} m\right]\right)^{-1}\widehat{\mathbb{E}}\left[\nabla_{\phi\vartheta} m\right],
\end{equation}
we obtain, taking probability limits at $(\overline\vartheta_0,\overline\phi_0)$,
\begin{equation}
\nabla_{\vartheta}\widehat s(\overline\vartheta_0)\overset{p}{\rightarrow}M_{\vartheta\vartheta}-M_{\vartheta\phi}\,M_{\phi\phi}^{-1}\,M_{\phi\vartheta}=:M_{\vartheta\vartheta\cdot\phi},
\end{equation}
where
\begin{equation}
M_{\vartheta\vartheta}={\mathbb{E}}_{\vartheta_0}\left[\nabla_{\vartheta\vartheta} m\left(\overline\vartheta_0,\overline\phi_0,y_{1:T}\right)\right],\qquad
M_{\phi\vartheta}=M_{\vartheta\phi}^\top.
\end{equation}
Combining the two limits,
\begin{equation}
\sqrt{N}\left(\widehat{\vartheta}^{\rm VI}-\overline\vartheta_0\right)=-M_{\vartheta\vartheta\cdot\phi}^{-1}\,\sqrt{N}\,\widehat{\mathbb{E}}\left[\widetilde s\left(y_{1:T}\right)\right]+o_p(1)\overset{d}{\rightarrow}{\cal{N}}\left(0,V^{\rm VI}\right),
\end{equation}
with
\begin{equation}
V^{\rm VI}=M_{\vartheta\vartheta\cdot\phi}^{-1}\,\widetilde\Omega\,M_{\vartheta\vartheta\cdot\phi}^{-\top}.
\end{equation}
This sandwich expression can be estimated consistently without having to evaluate the likelihood, as shown by \citet{westling2019beyond}.

\subsection{Expression of $V^{\rm IVI}$}

Since $b(\vartheta_0)=\overline\vartheta_0$ and $\widehat b$ is consistent for $b$, a first-order expansion of the identity $\widehat{b}(\widehat{\vartheta}^{\rm IVI})=\widehat{\vartheta}^{\rm VI}$ around $(\vartheta_0,b)$ gives
\begin{equation}
B\left(\widehat{\vartheta}^{\rm IVI}-\vartheta_0\right)=\left(\widehat{\vartheta}^{\rm VI}-\overline\vartheta_0\right)-\left(\widehat{b}(\vartheta_0)-b(\vartheta_0)\right)+o_p(N^{-\frac{1}{2}}),
\end{equation}
where 
\begin{equation}
B=\nabla_{\vartheta}b(\vartheta_0).
\end{equation}
The estimated binding function $\widehat b(\vartheta_0)$ is itself a profiled M-estimator, computed on simulated data $\widetilde y_{1:T}(\vartheta_0)$ drawn from ${\cal{P}}_{\vartheta_0}$. By the same argument as for $V^{\rm VI}$, applied to the simulated objective, its first-order expansion is
\begin{equation}
\widehat b(\vartheta_0)-b(\vartheta_0)=-M_{\vartheta\vartheta\cdot\phi}^{-1}\,\widetilde{\mathbb{E}}\left[\widetilde s\left(\widetilde y_{1:T}(\vartheta_0)\right)\right]+o_p(N^{-\frac{1}{2}}).
\end{equation}
Since the simulated draws are independent of the observed data and use $M$ independent draws per observation,
\begin{equation}
\sqrt{N}\left(\widehat b(\vartheta_0)-b(\vartheta_0)\right)\overset{d}{\rightarrow}{\cal{N}}\left(0,\tfrac{1}{M}\,V^{\rm VI}\right),
\end{equation}
independently of $\sqrt{N}(\widehat{\vartheta}^{\rm VI}-\overline\vartheta_0)$. Lastly, combining the two independent contributions,
\begin{equation}
\sqrt{N}\left(\widehat{\vartheta}^{\rm IVI}-\vartheta_0\right)=B^{-1}\left[\sqrt{N}\left(\widehat{\vartheta}^{\rm VI}-\overline\vartheta_0\right)-\sqrt{N}\left(\widehat b(\vartheta_0)-b(\vartheta_0)\right)\right]+o_p(1)\overset{d}{\rightarrow}{\cal{N}}\left(0,V^{\rm IVI}\right),
\end{equation}
with
\begin{equation}
V^{\rm IVI}=\left(1+\tfrac{1}{M}\right)B^{-1}\,V^{\rm VI}\,B^{-\top}.
\end{equation}
The factor $1+\tfrac{1}{M}$ is the usual simulation-noise inflation of indirect inference \citep{gourieroux1993indirect}. An empirical counterpart is
\begin{equation*}
\widehat V^{\rm IVI}=\left(1+\tfrac{1}{M}\right)\widehat B^{-1}\,\widehat V^{\rm VI}\,\widehat B^{-\top},
\end{equation*}
where $\widehat V^{\rm VI}$ is the \citet{westling2019beyond} estimator of $V^{\rm VI}$, and $$\widehat{B}=\nabla_{\vartheta}\widehat b\left(\widehat{\vartheta}^{\rm IVI}\right).$$
Computing $\widehat{B}$ requires evaluating the gradient of the binding function, which can be done without evaluating the likelihood (see Appendix \ref{app:grad}).

\section{Gradient Descent\label{app:grad}}

\subsection{Gradient of the binding function}

To implement gradient descent in (\ref{eq:grad}) we need $\nabla_{\vartheta}b(\vartheta)$. We present calculations for the population gradient. The calculations are analogous in the case of the empirical gradient. Recall that $b(\vartheta)$ maximizes the population ELBO, so it satisfies the first-order condition
\begin{equation*}
\nabla_{\widetilde\vartheta}\Big|_{\widetilde\vartheta=b(\vartheta)}\,\mathbb{E}_{\vartheta}\left[ m\left(\widetilde\vartheta,\overline\phi(\widetilde\vartheta,\vartheta),y_{1:T}\right)\right]=0,
\end{equation*}
where the outer expectation is taken over ${\cal{P}}_{\vartheta}$, and
$$\overline{\phi}(\widetilde\vartheta,\vartheta)
= \underset{\phi}{\mbox{argmax}}\ \mathbb{E}_{\vartheta}\big[\mathcal{E}_{\widetilde\vartheta,\phi}(y_{1:T})\big].$$ 
By the envelope theorem the inner dependence on $\overline\phi$ drops from the $\widetilde\vartheta$-gradient, so we may write the defining condition as $G(b(\vartheta),\vartheta)=0$ with
\begin{equation*}
G(\widetilde\vartheta,\vartheta)=\mathbb{E}_{\vartheta}\left[\nabla_{\widetilde\vartheta}\, m\left(\widetilde\vartheta,\overline\phi(\widetilde\vartheta,\vartheta),y_{1:T}\right)\right].
\end{equation*}
Differentiating the identity $G(b(\vartheta),\vartheta)=0$ totally in $\vartheta$ and applying the implicit function theorem,
\begin{equation*}
\nabla_{\vartheta}b(\vartheta)=-\left[\nabla_{\widetilde\vartheta}\,G\left(b(\vartheta),\vartheta\right)\right]^{-1}\nabla_{\vartheta}\,G\left(b(\vartheta),\vartheta\right).\label{eq:gradb}
\end{equation*}
The numerator $\nabla_{\vartheta}G\left(b(\vartheta),\vartheta\right)$ can be computed by simulations, using the reparameterization trick applied to $y_{1:T}$. Writing $y_{1:T}=h(\varepsilon,\vartheta)$ for a draw $\varepsilon$ from a $\vartheta$-free distribution, we have $G(\widetilde\vartheta,\vartheta)=\mathbb{E}_{\varepsilon}\left[\nabla_{\widetilde\vartheta}\, m\left(\widetilde\vartheta,\overline\phi(\widetilde\vartheta,\vartheta),h(\varepsilon,\vartheta)\right)\right]$, and since the distribution of $\varepsilon$ does not depend on $\vartheta$, differentiation passes inside the expectation,
\begin{equation*}
\nabla_{\vartheta}G\left(b(\vartheta),\vartheta\right)=\mathbb{E}_{\varepsilon}\left[\nabla_{\vartheta}\left\{\nabla_{\widetilde\vartheta}\, m\left(b(\vartheta),\overline\phi(b(\vartheta),\vartheta),h(\varepsilon,\vartheta)\right)\right\}\right].
\end{equation*}
The denominator $\nabla_{\widetilde\vartheta}G\left(b(\vartheta),\vartheta\right)$ is
\begin{align*}
\nabla_{\widetilde\vartheta}G\left(b(\vartheta),\vartheta\right)&=\mathbb{E}_{\vartheta}\left[\nabla_{\widetilde\vartheta\widetilde\vartheta}\, m\left(b(\vartheta),\overline\phi(b(\vartheta),\vartheta),y_{1:T}\right)\right]\\&+\mathbb{E}_{\vartheta}\left[\nabla_{\widetilde\vartheta\phi}\, m\left(b(\vartheta),\overline\phi(b(\vartheta),\vartheta),y_{1:T}\right)\right]\,\nabla_{\widetilde\vartheta}\overline\phi(b(\vartheta),\vartheta),
\end{align*}
where
\begin{equation*}
\nabla_{\widetilde\vartheta}\overline\phi(b(\vartheta),\vartheta)=-\left(\mathbb{E}_{\vartheta}\left[\nabla_{\phi\phi}\, m\left(b(\vartheta),\overline\phi(b(\vartheta),\vartheta),y_{1:T}\right)\right]\right)^{-1}\mathbb{E}_{\vartheta}\left[\nabla_{\phi\widetilde\vartheta}\, m\left(b(\vartheta),\overline\phi(b(\vartheta),\vartheta),y_{1:T}\right)\right].
\end{equation*}
Evaluating the denominator thus requires computing derivatives with the possible large parameter vector $\phi$. Note that although the envelope theorem removes the first-order $\phi$-sensitivity from $G$ itself, differentiating $G$ in $\widetilde\vartheta$ reintroduces a second-order term in $\phi$, which is the difficulty the fixed-$q$ approach in the next subsection avoids.

\subsection{Alternative binding function}

To present the approach we first conduct the analysis at the population level. Let $\overline\phi_0=\overline\phi(\overline\vartheta_0,\vartheta_0)$ denote the variational optimum obtained at the variational pseudo-true value. Holding this value fixed, we define the \emph{fixed-$q$ binding function}
\begin{equation*}
b^{\text{fixed-q}}(\vartheta)=\underset{\widetilde\vartheta}{\mbox{argmax}}\, \mathbb{E}_{\vartheta}\left[{\cal{E}}_{\widetilde\vartheta,\overline\phi_0}(y_{1:T})\right].
\end{equation*}
By construction, $b^{\text{fixed-q}}(\vartheta_0)=\overline{\vartheta}_0$, since at $\vartheta=\vartheta_0$ the simulated objective coincides with the population variational objective under ${\mathcal P}_{\vartheta_0}$ evaluated at the fixed $\overline\phi_0$, whose maximizer over $\widetilde\vartheta$ is $\overline\vartheta_0$ by definition. Hence, 
\begin{equation*}
\vartheta_0=\underset{\vartheta}{\mbox{argmin}}\,\left\|b^{\text{fixed-q}}(\vartheta)-\overline\vartheta_0\right\|^2.
\end{equation*}
The advantage of the fixed-$q$ approach is computational. The first-order condition defining $b^{\text{fixed-q}}$ is $$G^{\text{fixed-q}}(b^{\text{fixed-q}}(\vartheta),\vartheta)=0,$$ with
\begin{equation*}
G^{\text{fixed-q}}(\widetilde\vartheta,\vartheta)=\mathbb{E}_{\vartheta}\left[\nabla_{\widetilde\vartheta}\, m\left(\widetilde\vartheta,\overline\phi_0,y_{1:T}\right)\right],
\end{equation*}
and since $\overline\phi_0$ is now a constant rather than a function of $\widetilde\vartheta$, the gradient
\begin{equation*}
\nabla_{\vartheta}b^{\text{fixed-q}}(\vartheta)=-\left[\nabla_{\widetilde\vartheta}\,G^{\text{fixed-q}}\left(b^{\text{fixed-q}}(\vartheta),\vartheta\right)\right]^{-1}\nabla_{\vartheta}\,G^{\text{fixed-q}}\left(b^{\text{fixed-q}}(\vartheta),\vartheta\right)
\end{equation*}
involves the denominator
\begin{equation*}
\nabla_{\widetilde\vartheta}\,G^{\text{fixed-q}}\left(b^{\text{fixed-q}}(\vartheta),\vartheta\right)=\mathbb{E}_{\vartheta}\left[\nabla_{\widetilde\vartheta\widetilde\vartheta}\, m\left(\widetilde\vartheta,\overline\phi_0,y_{1:T}\right)\right]
\end{equation*}
that does not involve derivatives with respect to the variational parameter $\phi$. 

For estimation, we replace $\overline\phi_0$ by its consistent estimator
\begin{equation*}
\widehat\phi=\underset{\phi}{\mbox{argmax}}\,\widehat{\mathbb{E}}\left[{\cal{E}}_{\widehat\vartheta^{\rm VI},\phi}(y_{1:T})\right].
\end{equation*}
Holding $\widehat\phi$ fixed, we define the estimated fixed-$q$ binding function
\begin{equation*}
\widehat{b}^{\text{fixed-q}}(\vartheta)=\underset{\widetilde\vartheta}{\mbox{argmax}}\, \widetilde{\mathbb{E}}\left[{\cal{E}}_{\widetilde\vartheta,\widehat\phi}\big(\widetilde{y}_{1:T}(\vartheta)\big)\right],
\end{equation*}
where $\widetilde{y}_{1:T}(\vartheta)$ are simulated draws from ${\cal{P}}_{\vartheta}$ and $\widetilde{\mathbb{E}}$ denotes the sample mean over the $NM$ simulated draws. The fixed-$q$ indirect variational estimator is then obtained as
\begin{equation*}
\widehat{\vartheta}^{\text{IVI-fixed-q}}=\big[\widehat{b}^{\text{fixed-q}}\big]^{-1}\big(\widehat\vartheta^{\rm VI}\big),
\end{equation*}
computed by gradient descent or fixed-point iteration.

Because $b^{\text{fixed-q}}\neq b$ away from $\vartheta_0$, the two approaches need not produce identical estimates at each iteration, but they share the same fixed point $\vartheta_0$ and, under the one-to-one assumption, the same solution. The asymptotic distribution of the resulting estimator follows the same arguments as for $V^{\rm IVI}$, with the Jacobian $B$ replaced by $B^{\text{fixed-q}}=\nabla_{\vartheta}b^{\text{fixed-q}}(\vartheta_0)$.

\section{Technical Details\label{app:tech_details}}

\paragraph{Expression of $\Sigma_z$.}

We have
\[
\Sigma_z=
\begin{pmatrix}
V_1 & \rho V_1 & \rho^2 V_1 & \cdots & \rho^{T-1} V_1 \\
\rho V_1 & V_2 & \rho V_2 & \cdots & \rho^{T-2} V_2 \\
\rho^2 V_1 & \rho V_2 & V_3 & \cdots & \rho^{T-3} V_3 \\
\vdots & \vdots & \vdots & \ddots & \vdots \\
\rho^{T-1} V_1 & \rho^{T-2} V_2 & \rho^{T-3} V_3 & \cdots & V_T
\end{pmatrix},
\]
for (assuming $|\rho|<1$)
\[
V_t=\rho^{2(t-1)}\sigma_{z_1}^2+\sigma^2\frac{1-\rho^{2(t-1)}}{1-\rho^2},\qquad t=1,\dots,T.
\]

\paragraph{Expression of $\Sigma_z^{-1}$.}

We have
\begin{equation}
\Sigma_z^{-1}=
\begin{pmatrix}
\displaystyle \frac{1}{\sigma_{z_1}^2}+\frac{\rho^2}{\sigma^2} & \displaystyle -\frac{\rho}{\sigma^2} & 0 & \cdots & 0 \\[8pt]
\displaystyle -\frac{\rho}{\sigma^2} & \displaystyle \frac{1+\rho^2}{\sigma^2} & \displaystyle -\frac{\rho}{\sigma^2} & \cdots & 0 \\[6pt]
0 & \displaystyle -\frac{\rho}{\sigma^2} & \displaystyle \frac{1+\rho^2}{\sigma^2} & \ddots & \vdots \\[6pt]
\vdots & \ddots & \ddots & \ddots & \displaystyle -\frac{\rho}{\sigma^2} \\[6pt]
0 & \cdots & 0 & \displaystyle -\frac{\rho}{\sigma^2} & \displaystyle \frac{1}{\sigma^2}
\end{pmatrix}.\label{eq:sig_minus1}
\end{equation}

\paragraph{Penalized log-likelihood under mean-field.}

For a Gaussian variational density with parameters $\phi=(\mu_q,\Sigma_q)$, the KL divergence has the following analytical expression:
\begin{align*}
  &\mathrm{KL}\big[q_\phi(z_{1:T}\,|\, y_{1:T}) \,\|\, p_{\vartheta}(z_{1:T}\,|\, y_{1:T})\big]\\&=\frac{1}{2}\left(\log\frac{|\Sigma_{z|y}|}{|\Sigma_{q}|}-T+\text{tr}\left(\Sigma_{z|y}^{-1}\Sigma_q\right)+\left(\mu_{z|y}-\mu_q\right)^\top \Sigma_{z|y}^{-1}\left(\mu_{z|y}-\mu_q\right)\right),
\end{align*}
where $\mu_{z|y}$ and $\Sigma_{z|y}$ are given by (\ref{eq:muzy}) and (\ref{eq:sigzy}), respectively.

Suppose that $\mu_q$ is unrestricted. Further, following  the mean-field approach, suppose that $\Sigma_q$ is diagonal. Then, denoting as $\sigma_{q,t}^2$ the diagonal elements of $\Sigma_q$, and using the notation $\Omega^{\rm post}=\Sigma_{z|y}^{-1}$ for the posterior precision matrix, with diagonal elements $(\omega_{t}^{\rm post})^2$, we have
\begin{align*}
 &\underset{\mu_q,\Sigma_q}{\min}\,  \mathrm{KL}\big[q_\phi(z_{1:T}\,|\, y_{1:T}) \,\|\, p_{\vartheta}(z_{1:T}\,|\, y_{1:T})\big]\\
 &=\underset{\sigma_{q,1},...,\sigma_{q,T}}{\min}\, \frac{1}{2}\left(-\log|\Omega^{\rm post}|-\sum_{t=1}^T\log \sigma_{q,t}^2
 -T+\sum_{t=1}^T(\omega_{t}^{\rm post})^2\sigma_{q,t}^2\right)\\
 &=\frac{1}{2}\left(-\log|\Omega^{\rm post}|+\sum_{t=1}^T\log (\omega_{t}^{\rm post})^2
 \right)\\
 &=\frac{1}{2}\log \frac{|\mbox{diag}\, \Omega^{\rm post}|}{|\Omega^{\rm post}|},
\end{align*}
where $\mbox{diag}\, \Omega^{\rm post}$ is the diagonal of $\Omega^{\rm post}$.

Hence, the penalized log-likelihood in (\ref{eq:QL}) equals
\begin{align}
&\ {\mathcal{E}}^*_{\rho, \sigma, \sigma_e, \sigma_{z_1}}(y_{1:T}) \notag\\&= \underset{\text{log-likelihood}}{\underbrace{-\frac{T}{2} \log(2\pi) - \frac{1}{2} \log |\Sigma_z + \sigma_e^2 I_T| - \frac{1}{2} y_{1:T}^\top (\Sigma_z + \sigma_e^2 I_T)^{-1} y_{1:T}}}-\underset{\text{penalty}}{\underbrace{\frac{1}{2}\log \frac{|\mbox{diag}\, \Omega^{\rm post}|}{|\Omega^{\rm post}|}}}.\label{eq:QL_bench2}
\end{align}

Lastly, note that while the mean-field approach does not restrict the variational posterior mean, since $\mu_q=\mu_{z|y}$, it forces the variational posterior variances to be equal to the inverse diagonal elements of the precision matrix, $\sigma_{q,t}^2=(\omega_t^{\rm post})^{-2}$.

\section{Resource Use\label{app:resources}}

In Table~\ref{hockeystick_resource_table} we document resource use for two approaches: VI based on a Gaussian variational posterior, and Sequential Monte Carlo (SMC). We compare four sample sizes: $N\in\{1000, 30{,}000\}$ and $T\in\{6, 40\}$. In the left panel we report time to convergence, in the central panel we report time per iteration, and in the right panel we report GPU memory at the peak. 

The table shows that VI is faster and consumes less memory than SMC, given our implementation of the two methods and our convergence criterion. Moreover, the difference between the two methods is larger for bigger sample sizes.

\begin{table}[!t]
\centering
\caption{Resource Use for Variational Inference and Sequential Monte Carlo}
\label{hockeystick_resource_table}
\setlength{\tabcolsep}{2pt}%
\resizebox{\textwidth}{!}{%
\begin{tabular}{l@{\hspace{16pt}}cccc@{\hspace{16pt}}cccc@{\hspace{16pt}}cccc}
\toprule
 & \multicolumn{4}{c}{Time to convergence (min)} & \multicolumn{4}{c}{Time per iteration (ms)} & \multicolumn{4}{c}{Peak GPU memory (MB)} \\
\cmidrule(l{3pt}r{3pt}){2-5}\cmidrule(l{3pt}r{3pt}){6-9}\cmidrule(l{3pt}r{3pt}){10-13}
$N$ (thousands) & $1$ & $1$ & $30$ & $30$ & $1$ & $1$ & $30$ & $30$ & $1$ & $1$ & $30$ & $30$ \\
$T$ & $6$ & $40$ & $6$ & $40$ & $6$ & $40$ & $6$ & $40$ & $6$ & $40$ & $6$ & $40$ \\
\midrule
\textit{Variational Inference} &  &  &  &  &  &  &  &  &  &  &  &  \\
\quad $32$ hidden nodes & \makebox[2.5em][c]{0.9} & \makebox[2.5em][c]{5.2} & \makebox[2.5em][c]{0.5} & \makebox[2.5em][c]{2.4} & \makebox[2.5em][c]{6.6} & \makebox[2.5em][c]{30.7} & \makebox[2.5em][c]{6.9} & \makebox[2.5em][c]{39.7} & \makebox[2.5em][c]{18} & \makebox[2.5em][c]{39} & \makebox[2.5em][c]{53} & \makebox[2.5em][c]{695} \\
\quad $64$ hidden nodes & \makebox[2.5em][c]{1.2} & \makebox[2.5em][c]{5.4} & \makebox[2.5em][c]{0.7} & \makebox[2.5em][c]{2.9} & \makebox[2.5em][c]{6.3} & \makebox[2.5em][c]{28.5} & \makebox[2.5em][c]{7.0} & \makebox[2.5em][c]{40.1} & \makebox[2.5em][c]{18} & \makebox[2.5em][c]{40} & \makebox[2.5em][c]{56} & \makebox[2.5em][c]{699} \\
\addlinespace
\textit{Sequential Monte Carlo} &  &  &  &  &  &  &  &  &  &  &  &  \\
\quad $50$ particles & \makebox[2.5em][c]{3.2} & \makebox[2.5em][c]{8.4} & \makebox[2.5em][c]{1.3} & \makebox[2.5em][c]{4.5} & \makebox[2.5em][c]{9.0} & \makebox[2.5em][c]{49.1} & \makebox[2.5em][c]{9.2} & \makebox[2.5em][c]{55.5} & \makebox[2.5em][c]{5} & \makebox[2.5em][c]{19} & \makebox[2.5em][c]{166} & \makebox[2.5em][c]{585} \\
\quad $200$ particles & \makebox[2.5em][c]{--} & \makebox[2.5em][c]{6.8} & \makebox[2.5em][c]{3.6} & \makebox[2.5em][c]{12.2} & \makebox[2.5em][c]{9.2} & \makebox[2.5em][c]{52.5} & \makebox[2.5em][c]{25.1} & \makebox[2.5em][c]{169.8} & \makebox[2.5em][c]{21} & \makebox[2.5em][c]{74} & \makebox[2.5em][c]{645} & \makebox[2.5em][c]{2221} \\
\bottomrule
\end{tabular}%
}
\\ \vspace{0.2cm} \figurenote{Resource cost of estimating the nonlinear DGP by variational inference (VI) and by sequential Monte Carlo (SMC), across four data sizes $(N,T)$. VI is shown for two hidden widths of the neural network and SMC for two particle counts. The convergence criterion is reached when the difference in the (running-averaged) parameter vector stays below $0.02$. Runs share a common seed and a budget of $25{,}000$ epochs. NVIDIA L40S GPU.}
\end{table}

\section{Computational Appendix}\label{app:comp}

This appendix records the estimation choices behind the five IVI results reported in the paper: the AR(1)-normal simulation (Section \ref{sec:benchmark}), the nonlinear model simulation at $T = 6$ and at $T = 40$ (Section \ref{sec:nonlinear-model}), the permanent heterogeneity simulation (Subsection \ref{sec:hetero-model}), and the model estimated on the PSID panel (Section \ref{sec:PSID}). Table~\ref{tab:computational-appendix} summarizes the settings.

\paragraph{Inner VI problem.}

Every fit maximizes the reparameterized ELBO with \textit{AdamW} at learning rate $10^{-2}$ on the full panel (no mini-batching), with gradient-norm clipping at $5$ and up to ten seed-offset retries on early NaN blow-ups. The posterior is a one-hidden-layer network of a given width $d$, mapping $y_{1:T}$ to the mean vector and Cholesky factor of a Gaussian posterior over the per-individual latents: the $T$ states $z_{1:T}$, plus the heterogeneity $a$ in the two heterogeneity specifications (latent dimension $T+1$). The AR(1) column deliberately uses the \emph{mean-field} (diagonal) family with $d = 32$, while all other columns use the full joint-normal Cholesky family with $d = 64$. Table~\ref{tab:computational-appendix} reports the number of parameters in the variational posterior, showing how the $T=40$ panel drives the parameter count to more than $58{,}000$, against roughly $2{,}000$ at $T=6$.

\paragraph{Outer IVI loop.}

The outer loop iterates on the \emph{binding function}, $b(\vartheta)$. The VI estimator is applied to a panel simulated at $\vartheta$, and the update is calculated as in equation (\ref{eq:FP}), with a damping factor of $\kappa=0.6$. The number of iterations is fixed in advance (no early stopping) and scales with the difficulty of the problem: 10 iterations suffice at $T = 6$, the nonlinear model uses 25 ($T=6$) and 50 ($T=40$), and the PSID fit uses 50. Each outer iteration re-fits the posterior from a cold start on the simulated panel.

\begin{table}[ht]
\centering
\footnotesize
\setlength{\tabcolsep}{2.5pt}
\caption{Estimation Settings for the Results}
\label{tab:computational-appendix}
\begin{tabular}{lccccc}
\toprule
 & {AR(1)} & {Nonlinear ($T{=}6$)} & {Nonlinear ($T{=}40$)} & {Heterogeneity} & {PSID} \\
\midrule
\multicolumn{6}{l}{\textit{Data and model}} \\
Panel $N \times T$ & $30{,}000 \times 6$ & $30{,}000 \times 6$ & $30{,}000 \times 40$ & $30{,}000 \times 6$ & $741 \times 10$ \\
Data & simulated & simulated & simulated & simulated & 1980--89 \\
Free model parameters & 8 & 12 & 12 & 12 & 15 \\
\midrule
\multicolumn{6}{l}{\textit{Variational posterior}} \\
Family & mean-field & joint normal & joint normal & joint normal $+\,a$ & joint normal $+\,a$ \\
Hidden width $d$ & 32 & 64 & 64 & 64 & 64 \\
Latent dimension & 6 & 6 & 40 & 7 & 11 \\
Var. posterior parameters & 620 & 2\,203 & 58\,524 & 2\,723 & 5\,709 \\
\midrule
\multicolumn{6}{l}{\textit{Inner VI (ELBO maximization)}} \\
ELBO draws  & 1 & 1 & 1 & 1 & 40 \\
Learning rate & $10^{-2}$ & $10^{-2}$ & $10^{-2}$ & $10^{-2}$ & $10^{-2}$ \\
Epochs, fit on data & 10\,000 & 20\,000 & 20\,000 & 20\,000 & 20\,000 \\
Epochs, per outer iter. & 8\,000 & 16\,000 & 16\,000 & 16\,000 & 16\,000 \\
\midrule
\multicolumn{6}{l}{\textit{Outer IVI loop}} \\
Outer iterations & 10 & 25 & 50 & 10 & 50 \\
Damping $\kappa$ & 0.6 & 0.6 & 0.6 & 0.6 & 0.6 \\
\bottomrule
\end{tabular}
\\ \vspace{0.2cm} \figurenote{This table summarizes the estimation settings for the AR(1)-normal simulation (Section \ref{sec:benchmark}), the nonlinear model simulation at $T = 6$ and at $T = 40$ (Section \ref{sec:nonlinear-model}), the permanent heterogeneity simulation (Subsection \ref{sec:hetero-model}), and the model estimated on the PSID panel (Section \ref{sec:PSID}).}
\end{table}

\paragraph{Random numbers.}

To make the binding function $b(\vartheta)$ a deterministic function of $\vartheta$, we fix every source of randomness across outer iterations: (i) the simulator shocks (draws from $z_{1:T}$, $a$, and the transitory shocks) are drawn once and reused; (ii) the reparameterization draws of the ELBO are drawn once (seed 12345) and reused at every epoch of every inner fit; (iii) the variational posterior is re-initialized from the same seed with a freshly constructed optimizer at every outer iteration.

\paragraph{ELBO draws.}

The simulated panels are large ($N = 30{,}000$), so a single reparameterization draw per gradient step already gives an accurate full-batch gradient. The PSID panel has $N = 741$: at that size the single-draw gradient is too noisy for the binding map to be estimated precisely, so the data fit averages $40$ fixed draws per ELBO evaluation (raising the effective draw count in place of the panel size) and uses as simulated panel size the same number of individuals as in the data.

\paragraph{Standard errors and reference likelihoods.}

On PSID, standard errors come from a nonparametric bootstrap: $B = 100$ replicates resample the 741 households with replacement and re-run the \emph{full} IVI pipeline (VI plus 20 fixed-point iterations at $40$ ELBO draws). The seeds are different across replicates. 

\section{Extension: Serially correlated shocks\label{app:MA1}}

To reflect the structure of the model with MA(1) transitory shocks presented in Subsection \ref{subsec_MA1}, one can specify a variational posterior that conditions each latent state on its lag \(z_{t-1}\), and on a residual \(r_t\) that captures relevant past information:
\begin{align}
    q(z_{1:T} \,|\, y_{1:T}) 
    = q(z_1 \,|\, y_{1:T}) 
      \prod_{t=2}^{T} q(z_t \,|\, z_{t-1}, y_{t:T}, r_{t-1}), 
    \label{eq:structured-q}
\end{align}
where the residual is defined recursively as
\begin{equation}
r_t = y_t - z_t - \zeta \, r_{t-1}, 
\quad r_1 = y_1 - z_1.\label{eq:structured-q-rt}
\end{equation}
Including \(r_{t-1}\) in the conditioning set enables the variational posterior to adapt to the dependence induced by the MA(1) structure and any non-Gaussianity in \(\varepsilon_t\), while keeping the approximation computationally tractable.

\section{Two Alternatives to Gaussian Variational Inference}\label{sec:extensions}

Our results based on simulated data indicate that, while Gaussian variational approximations tend to recover the conditional mean and variance quite well, the method fails at capturing the kurtosis of transitory shocks. Our approach to improve over VI is to rely on indirect inference and IVI. However, other strategies have been proposed in the literature to improve the accuracy of VI. Here we briefly review two of these strategies.

\subsection{Normalizing Flows}

Originally proposed by \citet{rezende2015variational}, a normalizing flow is an invertible transformation $h$ that maps a simple baseline distribution (e.g. a standard Gaussian) into a complex distribution. By incorporating flows into variational inference, we can start with a Gaussian $q_0(z)$ as baseline density, 
and define a sequence of transformations \( z_K = h_K \circ \cdots \circ h_1(z_0) \). The resulting density \( q_K(z_K) \) is computed using the change-of-variables formula. By choosing transformations \( h_k \) with tractable Jacobians and expressive functional forms, the variational family can capture complex density shapes, including skewness, heavy tails, and multimodality. One practical advantage is that flows maintain computational tractability (the transformations are chosen so that Jacobian determinants are easy to compute), so the ELBO with a flow-based $q_K(z_K)$ can still be optimized efficiently. 

To capture non-Gaussian features of the posterior distribution such as skewness and excess kurtosis, consider a one-dimensional transformation of a standard normal variable based on the sinh--arcsinh family introduced in Section \ref{sec:nonlinear-model}. This family allows one to preserve the tractability and reparameterization benefits of the Gaussian while introducing controlled deviations from normality. Let $\widetilde{z} \sim \mathcal{N}(0, 1)$ be a standard normal random variable. Next, define the transformed latent variable $z$ as
\begin{align}
z = T_{\nu}(\widetilde{z})=\mu + \sigma \cdot \sinh\left( \frac{ \operatorname{arcsinh}(\widetilde{z}) + \epsilon }{ \delta } \right),
\label{eq:sas-transform}
\end{align}
where $\mu \in \mathbb{R}$ and $\sigma > 0$ control the location and scale, $\delta > 0$ governs the kurtosis, $\epsilon \in \mathbb{R}$ introduces skewness, and $\nu = (\mu, \sigma, \epsilon, \delta)$. When $\delta = 1$ and $\epsilon = 0$, the transformation reduces to the identity and $z \sim \mathcal{N}(\mu, \sigma^2)$. The reparameterization trick applies directly, as $z$ is obtained by a differentiable transformation of a baseline normal variable. Gradients of the ELBO with respect to $\nu$ can then be estimated via differentiation. All terms are available in closed form, and the gradients are computed efficiently. At the same time, the sinh--arcsinh family has a specific, potentially restrictive functional form. The transformation is applied element-wise and does not model correlations across latent dimensions.

\subsection{Importance-Weighted Variational Inference}

Another approach is to use the fitted Gaussian variational posterior as a proposal density for \textit{importance sampling}. This strategy is referred to as Importance-Weighted Autoencoders (IWAE) (\citealp{wu2016quantitative}, \citealp{cremer2017reinterpreting}, \citealp{kim2020semi}). It exploits multiple draws from the variational posterior to tighten the ELBO, thereby reducing the gap with the true log-likelihood.

Formally, let \( q_\phi(z_{1:T} \,|\, y_{1:T}) \) denote a variational posterior family with parameter \( \phi \), and suppose we draw \( K \) independent values \( \{z^{(k)}\}_{k=1}^K \sim q_\phi(z \,|\, y) \). The importance-weighted ELBO is defined as
\begin{align*}
\mathcal{E}^{(K)}_{\vartheta,{\phi}}(y_{1:T}) = \mathbb{E}_{z^{(1)}, \dots, z^{(K)} \sim q_\phi(z_{1:T} \,|\, y_{1:T})} \left[ 
\log \left( \frac{1}{K} \sum_{k=1}^K \frac{f_\theta(z^{(k)}_{1:T})  \psi_\gamma(y_{1:T} - z^{(k)}_{1:T})}{q_\phi(z^{(k)}_{1:T}\,|\, y_{1:T})} \right) \right],
\end{align*}
where \( K \geq 1 \) controls the tightness of the bound. For \( K = 1 \), this recovers the standard ELBO; larger \( K \) values yield tighter bounds that approach the true log-likelihood from below. Indeed, it follows from Jensen's inequality that
$$\underset{\text{ELBO}}{\underbrace{\mathcal{E}_{\vartheta,{\phi}}(y_{1:T})}}=\mathcal{E}^{(1)}_{\vartheta,{\phi}}(y_{1:T})\leq ...\leq \mathcal{E}^{(K)}_{\vartheta,{\phi}}(y_{1:T})\leq \mathcal{E}^{(K+1)}_{\vartheta,{\phi}}(y_{1:T})\leq ...\leq \underset{\text{log-likelihood}}{\underbrace{\mathcal{L}_{\vartheta}(y_{1:T})}}.$$

\paragraph{Diagnostics. }

The normalized importance weights also allow the researcher to compute two diagnostics at essentially no additional cost. The \emph{effective sample size} (ESS) measures how many independent draws from the true posterior are effectively represented by the draws used in importance sampling. Formally, given the normalized importance weights
\[
\widetilde{w}_k = \frac{w_k}{\sum_{j=1}^K w_j},\quad \text{ with }\quad w_k = \frac{f_\theta(z^{(k)}_{1:T})  \psi_\gamma(y_{1:T} - z^{(k)}_{1:T})}{q_\phi(z^{(k)}_{1:T}\,|\, y_{1:T})},
\]
where $z^{(k)}$ are independent draws from the variational posterior \(q_{\phi}\), the ESS is defined as
\[
\mathrm{ESS} =  \frac{1}{K \sum_{k=1}^K \widetilde{w}_k^2}.
\]
This quantity lies between zero and one. An ESS close to one indicates that the variational posterior aligns well with the true posterior, since all samples contribute evenly to the importance-weighted estimate. Conversely, an ESS close to zero indicates severe degeneracy, with nearly all weights concentrated on a single draw, implying that the approximation \(q\) fails to cover the posterior adequately. 

Another useful diagnostic for assessing the quality of the variational approximation is the \emph{ELBO gap} between the evidence lower bound and the log-likelihood,
$$ {\cal{G}}_{\vartheta,\phi}(y_{1:T})={\cal{L}}_{\vartheta}(y_{1:T})-\mathcal{E}_{\vartheta,\phi}(y_{1:T}).$$
Computing the ELBO gap ${\cal{G}}_{\vartheta,\phi}(y_{1:T})$ exactly is often infeasible, since it requires evaluating the log-likelihood ${\cal{L}}_{\vartheta}(y_{1:T})$. However, unbiased Monte Carlo estimators of ${\cal{L}}_{\vartheta}(y_{1:T})$ can be constructed using importance sampling, using the same draws as in the IWAE approach. These estimators are considerably more expensive to compute than the standard ELBO. As a result, the ELBO gap is rarely used as an objective for optimization. Instead, it is most useful as an \emph{ex-post} diagnostic.

\end{document}